\documentclass[aps, prx, nofootinbib, preprintnumbers, twocolumn,10pt]{revtex4-2}

\usepackage[utf8]{inputenc}
\usepackage{amssymb}
\usepackage{mathtools}

\usepackage{epsfig}
\usepackage{epstopdf}
\usepackage{latexsym}
\usepackage{enumerate} 
\usepackage{graphicx}
\usepackage{placeins}
\usepackage{floatrow}
\usepackage{subfig}
\usepackage{caption}
\usepackage{booktabs}
\usepackage{bbm}
\usepackage{physics}
\usepackage{tensor}
\usepackage[compat=1.1.0]{tikz-feynman}
\usepackage{tikz}
\usetikzlibrary{shapes.geometric}
\usetikzlibrary{matrix}
\usetikzlibrary{decorations.markings,calc,shapes,decorations.pathmorphing,patterns,decorations.pathreplacing,arrows.meta}
\usetikzlibrary{positioning}

\tikzset{
wl/.style={scalar, thick},
w/.style={thick, decoration={markings, mark=at position 0.55 with {\arrow{Stealth}}},
    postaction={decorate}},
wbar/.style={thick, decoration={markings, mark=at position 0.55 with {\arrow{Stealth[reversed]}}},
    postaction={decorate}},
sigma/.style={thick, photon},
source/.style={shape=circle,
      draw=black,
      fill=white,
      scale=0.65,
      path picture={
        \draw[black]
          ([shift={(-3pt,-3pt)}]path picture bounding box.center)
            -- ([shift={(3pt,3pt)}]path picture bounding box.center);
        \draw[black]
          ([shift={(3pt,-3pt)}]path picture bounding box.center)
            -- ([shift={(-3pt,3pt)}]path picture bounding box.center);
      }
      }
}

\usepackage{xcolor}

\definecolor{darkblue}{rgb}{0.15,0.35,0.65}
\definecolor{reddish}{rgb}{0.65, 0.2, 0.2}
\definecolor{darkgreen}{RGB}{50,150,0}

\usepackage[linktocpage=true]{hyperref}
\hypersetup{
colorlinks=true,
citecolor=darkblue,
linkcolor=darkblue,
urlcolor=reddish,
pdfauthor={},
pdftitle={},
pdfsubject={}
}

\renewcommand{\a}{\mathfrak a}
\renewcommand{\b}{\mathfrak b}
\renewcommand{\c}{\mathfrak c}
\renewcommand{\d}{\mathfrak d}

\newcommand\Tstrut{\rule{0pt}{4ex}}         

\begin{document}

\title{Naturalness of vanishing black-hole tides}

\author{Julio Parra-Martinez}
\email{julio@ihes.fr}
\author{Alessandro Podo}
\email{podo@ihes.fr}
\affiliation{Institut des Hautes \'Etudes Scientifiques, 91440 Bures-sur-Yvette, France}

\begin{abstract}
We provide a symmetry argument for the vanishing and non-renormalization of static Love numbers for spherically symmetric black holes at full nonlinear order in four-dimensional General Relativity. The symmetry is realized both in full GR and in the worldline EFT, allowing for a unified treatment and proving both vanishing and non-renormalization to all orders. This closes some loop-holes in previous arguments that neglected nonlinearities in the worldline EFT, and extends previous vanishing results to all nonlinear static tides. 
When extended to higher-dimensional gravity, these arguments also explain the pattern of vanishing and running static Love numbers of electric and tensor type, and predict new results at the nonlinear order. We also apply our findings to the tidal response of shift-symmetric scalar fields, predicting the vanishing of even order nonlinear static Love numbers and unifying these statements with the no-hair theorem (and its violations).
\end{abstract}

\maketitle

{\bf \em Introduction.--} Arguably, the deepest modern puzzles in physics are naturalness problems. These are often thought to be exclusive to the quantum realm, but the classical world is not devoid of them. In particular, in the classical theory of General Relativity (GR) four-dimensional black holes seem to have a most mysterious and unnatural feature: they do not have an induced tidal response under time-independent external perturbations. In other words, black holes do not polarize under external DC potentials. Evidence for this fact has been mounting over the years, going back to an observation by Damour~\cite{Damour:1982wm}, together with recent explicit computations of linear responses~\cite{Fang:2005qq,Damour:2009vw,Binnington:2009bb,Damour:2009va,Kol:2011vg,Hui:2020xxx,Ivanov:2022hlo,Hadad:2024lsf}; and there is even evidence that a subset of nonlinear tides vanish~\cite{Poisson:2020vap,Poisson:2021yau,Riva:2023rcm,Iteanu:2024dvx,Kehagias:2024rtz,Combaluzier-Szteinsznaider:2024sgb,Gounis:2024hcm}.

The realization that this is a naturalness problem is most stark when parameterizing tidal responses using a point-particle Effective Field Theory (EFT)~\cite{Goldberger:2004jt, Porto:2005ac}, in which they correspond to (Wilson) coefficients of higher-dimensional operators, also known as Love numbers. Absent any symmetry in the EFT that allows us to set these coefficients to zero~\cite{tHooft:1979rat}, it would seem that black holes are highly unnatural and fine-tuned systems~\cite{Damour:1982wm,Porto:2016zng}. Various symmetries have been proposed to explain the vanishing of the linear tidal coefficients~\cite{Charalambous:2021kcz,Hui:2021vcv,Charalambous:2022rre,Hui:2022vbh,Berens:2022ebl,BenAchour:2022uqo,Sharma:2024hlz}, and of axisymmetric nonlinear tides of electric type~\cite{Kehagias:2024rtz,Combaluzier-Szteinsznaider:2024sgb,Gounis:2024hcm}. However, it is unclear how some of these symmetries act in the EFT (see, however,~\cite{Berens:2025okm}) and how they generalize for generic nonlinear perturbations, thus precluding a fully satisfactory and uniform explanation.

In this paper we provide a solution to this naturalness problem, and greatly extend previous results, by identifying an accidental symmetry of static GR which predicts the \emph{vanishing of the static tidal responses, or Love numbers, of Schwarzschild black holes at the full nonlinear level}.
In other words, we find that:
\begin{quote}
\emph{Black holes do not acquire induced field multipole moments, no matter how strong an external time-independent field is applied to them.} 
\end{quote}
The symmetry implying this remarkable feature is not entirely new. It has been long known that there is an ${\rm SL}(2,\mathbb{R})$ symmetry relating gravitational solutions with a timelike Killing vector, discussed by Geroch~\cite{Geroch:1970nt} extending previous works of Buchdahl, Ehlers, and Ernst~\cite{Buchdahl:1954,Ehlers:1959aug, Ernst:1967wx} (see~\cite{Stephani:2003tm} for a textbook treatment). While this is usually understood as a solution-generating symmetry, here we point out that this is in fact a fully off-shell accidental symmetry of GR in the static sector, valid at the level of the action.\footnote{For consistency with the standard nomenclature in the literature on tides, we will refer to the collection of time-independent gravitational configurations as the static sector. This includes both static and stationary spacetimes. The latter, in particular, are crucial to define and probe tidal responses of magnetic type.} These symmetries are spontaneously broken by the parameters in particular solutions, or explicitly by the EFT coefficients. This enables us to show the vanishing of the tides using a conventional spurion argument. In fact only a $\mathbb{Z}_2$ subgroup of ${\rm SL}(2,\mathbb{R})$ is necessary for the argument.

Our results apply both for electric and magnetic responses, thus unifying and generalizing previous partial results. Furthermore, the arguments extend to a series of electric and tensor tides in higher dimensions ($D>4$), as well as for black holes in theories with shift-symmetric scalars, showing the wide applicability of our method. We focus on spherically-symmetric solutions, leaving the analysis of spinning black holes for further work.

{\bf \em Nonlinear accidental symmetries of static General Relativity.--} Our goal is to study generic time-independent gravitational perturbations. A useful parametrization for this class of metrics is Kaluza-Klein-like~\cite{Ehlers:1959aug} (see also~\cite{Kol:2011vg}):
\begin{equation}\label{eq:temporalKK}
ds^2 = - e^{2\phi} \left(dt - A_i dx^i \right)^2 + e^{-2\phi} \gamma_{ij} dx^i dx^j.
\end{equation}
In what follows, three-dimensional indices $i,j,k, \dots$ are raised and lowered with the three-dimensional metric $\gamma_{ij}$ and its inverse $\gamma^{ij}$. The Ricci scalar $R[\gamma]$ refers to that associated to the metric $\gamma_{ij}$, while $F_{ij}=\partial_iA_j-\partial_jA_i$.

In this field basis, the Einstein-Hilbert action is
\begin{align}
        \begin{split}
        S_{\rm EH} &= - \dfrac{1}{16\pi G} \int d^{4}{x} \sqrt{\gamma} \; \Big(-R[\gamma] \\  &\hspace{1cm}  + 2 (\partial_i\phi)^2 - \dfrac{1}{4} e^{4\phi} F_{ij}^2 +{\cal O} (\partial_t^2)\Big) ,
\end{split}
\end{align}
where we have dropped terms with time-derivatives, since we focus on the static limit.

The graviphoton $A_i$ can be dualized to a pseudoscalar ``axion'' $a$ modulo time derivatives (see Appendix~\ref{app:dualization}) 
\begin{equation}
 (\nabla \times A)_i = \dfrac{1}{2}\epsilon_{ i}^{\phantom{a}jk} F_{jk} = e^{-4\phi} \partial_i a  + {\cal O} (\partial_t) \, .
\end{equation}
Introducing the complex scalar
\begin{equation}
z= a +i e^{2\phi},
\end{equation}
which lives in the upper-half plane ${\rm Im}\,z>0$, the action takes the form
\begin{equation}
\label{eq:zaction}
S_{\rm EH} = - \dfrac{1}{16\pi G} \int d^{4}{x} \sqrt{\gamma} \; \Big(-R[\gamma] + \dfrac{1}{2} \dfrac{\partial_i z \,\partial^i\bar{z}}{({\rm Im}\,z)^2}  +{\cal O} (\partial_t^2) \Big).
\end{equation}
This action makes manifest that the static sector of four dimensional GR is invariant under an accidental ${\rm (P)SL}(2,\mathbb{R})$ symmetry
\begin{equation}
z \to \dfrac{\a z+\b }{\c z+\d }, \quad {\rm with } \quad \a \d-\b \c =1,
\end{equation}
and $\a,\b,\c,\d \in \mathbb{R}$.
The transformations
\begin{equation}
S: z \to -\dfrac{1}{z}, \quad {\rm and} \quad T: z \to z+ \b,
\end{equation}
generate the whole ${\rm SL}(2,\mathbb{R})$. The $S$ transformation is highly nonlinear, but when $a=0$ it simply corresponds to the $\phi \to -\phi$ symmetry discussed in~\cite{Kol:2011vg} (see also~\cite{Ivanov:2022hlo,santoni2024black,Hadad:2024lsf,Diedrichs:2025vhv}). The $T$ transformation is simply the shift symmetry of the axion $a\to a+\b$. The ${\rm SL}(2,\mathbb{R})$ symmetry is realized nonlinearly, i.e., it is spontaneously broken down to an ${\rm SO}(2)$ subgroup. Indeed the action for $z$ can be identified as the nonlinear sigma model (NLSM) for the coset space ${\rm SL}(2,\mathbb{R})/{\rm SO}(2)$. 
The metric corresponds to the hyperbolic plane and the field variable $z$ realizes the Poincar\'e half-plane model of hyperbolic geometry. In addition, the theory is invariant under the parity transformation
\begin{equation}
P: z \to -\bar{z},
\end{equation}
which corresponds to $a(x)\to -a(x)$.

It is useful to map the upper-half plane to the unit disk by the field redefinition
\begin{equation}
w = \dfrac{z-i}{z+i} = \phi - i \, \dfrac{a}{2} + \cdots \,.
\end{equation}
In this parametrization the unbroken ${\rm SO}(2)$ symmetry becomes manifest as a rotation around the origin \begin{equation}\label{so2}
    w \to e^{i \theta} w\,.
\end{equation} 
The $S$ transformation corresponds to $\theta=\pi$ and generates a $\mathbb{Z}_2^S$ which simply sends $w\to -w$. Parity acts as $w\to \overline{w}$.
We refer to Appendix~\ref{app:dualization} for a more detailed discussion and for the explicit form of the action in terms of $w$.

{\bf \em Symmetry breaking and spurion analysis.--} The Minkowski background is obtained for $a=\phi=0$, that is, $w=0$ (or $z=i$), while $\gamma_{ij}=\delta_{ij}$. In such background the ${\rm SO}(2)$ symmetry in Eq.~\eqref{so2} remains unbroken. On the other hand, the Schwarzschild background in isotropic coordinates (which we use throughout this paper) is given by
\begin{equation}
ds^2 = -\left(\dfrac{1-X}{1+X} \right)^2 dt^2 + \Big(1+X\Big)^{4} \, \delta_{ij} \, dx^i dx^j,
\end{equation}
where 
\begin{equation}\label{eq:Xiso}
X \equiv \dfrac{1}{4}\left(\dfrac{R_S}{r}\right) = \dfrac{GM}{2r} \qquad r = \sqrt{\delta_{ij}x^ix^j}\,.
\end{equation}
In terms of $w$ and $\gamma$ this is 
\begin{equation}
w_{\rm Sch} = -\dfrac{2X}{1+X^2}
, \quad \gamma_{ij}\Big\vert_{\rm Sch} =(1-X^2)^2 \delta_{ij}\,,
\label{eq:wgSchw}
\end{equation}
which further spontaneously breaks the ${\rm SO}(2)$ in \eqref{so2}.

The $\mathbb{Z}_2^S$ symmetry $w\to -w$, on the other hand, can be restored as a spurionic symmetry if we assign a transformation rule $M\to -M$ to the mass.
The Schwarzschild solution is therefore invariant under ${w\to -w}$ if we transform ${X\to -X}$. Since $X$ is proportional to $M$, this is just the transformation rule ${M \to -M}$.

The question of the tides concerns the static non-linear perturbations around a Schwarzschild black hole in the presence of a source. Since both the Einstein-Hilbert action and the background solution are symmetric under $\mathbb{Z}_2^S$ if we transform ${M\to -M}$, this symmetry will also be present in the nonlinear action for such perturbations. The action of the symmetry on the perturbations will have a more complicated form than that on $z$, but the explicit form will not be important for our arguments.
Indeed, consider studying the (nonlinear) equations for the perturbations of the space-time corresponding to some given compact object. Since the action for the perturbations is $\mathbb{Z}_2^S$ symmetric (with $M\to-M$), so are the equations of motion. Therefore, the only way the solution can break the symmetry is if the boundary conditions break the symmetry.\footnote{We are grateful to Austin Joyce and Luca Santoni for a question on the role of boundary conditions, that allowed us to improve the presentation.} The equations are of second order and thus require two boundary conditions. For the source term, $g_{\rm src}\sim r^{\ell}$ at large $r$, we can always assign transformation rules consistent with the symmetries (e.g. $w_{\rm src}\to-w_{\rm src}$). The only source of symmetry breaking can therefore come from the second boundary condition imposed at some scale $R_\star$. If there is no dimensionful scale in the problem other then the Schwarzschild radius $R_S$, the boundary condition will specify the solutions as a function of $R_S/R_\star$. Since $R_S \to -R_S$ this will break the symmetry unless only even terms are present. For black holes, the condition we impose is regularity at the horizon, which corresponds to a numerical constant at $R_\star=R_S$. (Notice that $R_\star$ does not transform under the symmetry.) The regular solution is therefore symmetric and we can use the symmetry to constrain the form of the tail. In the case of stars this argument does not apply because the relevant boundary conditions depend on new scales and break the symmetry.

As we will see next, this property together with the absence of nonlinear contributions in the EFT matching computation, implies the vanishing of Love numbers for black holes, and the absence of running.

{\bf {\em Selection rules: vanishing and non-renormalization of static Love numbers of Schwarzschild black holes.--}} We can now use these accidental symmetries to derive selection rules, and argue for the vanishing and non-renormalization of the nonlinear static tides of Schwarzschild black holes. As we will become clear later (see eq.~\eqref{eq:EBw}) the complex field $w$ allows to conveniently capture all the electric and magnetic responses, encoded by its real and imaginary parts respectively.

In the presence of a source $w_{\rm src} \sim r^\ell$, the dimensionless static Love numbers $k_\ell$ are usually defined as the coefficient of the $(2\ell+1)$-th term in the Post-Newtonian (PN) expansion of the linear response (or ``tail'') of the field
\begin{equation}
w_{\rm tail} \sim \,w_{\rm src} \left(1+\dots + k_\ell \, X^{2\ell +1} + \dots  \right),
\end{equation}
where we remind the reader that $X\sim R_S/r$.
Now we use the symmetry to argue that $k_\ell$ must be zero. We begin by noting that both the source and the response transform equally under the spurionic $\mathbb{Z}_2$ symmetry:
\begin{equation} \label{eq:z2wbartail}
    w_{\rm src}\to - w_{\rm src}\,, \quad w_{\rm tail} \to - w_{\rm tail}\,.
\end{equation}
Nevertheless, the symmetry also requires $M\to - M$, which takes $X\to - X$, so the tail transforms as
\begin{align}
    &w_{\rm src} \left(1+\dots + k_\ell \, X^{2\ell +1} + \dots  \right) \\
    &\hspace{1cm}\to - w_{\rm src} \left(1+\dots - k_\ell \, X^{2\ell +1} + \dots  \right)\; . \nonumber 
\end{align}
Thus, the response is only compatible with the symmetry in Eq.~\eqref{eq:z2wbartail} if $k_\ell$=0, and hence the linear static Love numbers, vanish. In fact our argument shows more generally that all terms in the linear response of $w$ which are odd in $X$ must be zero. We have verified this explicitly by solving for the linear perturbations in isotropic coordinates and imposing regularity at the horizon~\cite{LongTidesPaper}.\footnote{Our computation can be phrased in terms of curvature perturbations of the Weyl tensor, so regularity at the Schwarzschild horizon is a necessary condition.} For linear tides of electric type this is also manifest in the results of~\cite{Kol:2011vg,Diedrichs:2025vhv}. 

Note that the symmetry we used for the argument does not only amount to $\phi \to -\phi$, since this is not a symmetry of the fully nonlinear static gravitational action. The inversion symmetry $S$ is, instead, a nonlinear transformation that involves both $\phi$ and $a$. It reduces to the Buchdahl $\mathbb{Z}_2$ transformation~\cite{Buchdahl:1954} when $a=0$, such as on the Schwarzschild background. Thus, in the case of purely electric tidal response, the $\phi \to -\phi$ symmetry is sufficient for our arguments. The treatment of magnetic and mixed tides, however, requires the use of the full $S$-type transformation.

Let us comment on an additional subtlety. Generically, the response coefficients could have contained logarithmic terms $k_\ell= k_\ell^0 + \beta_\ell \ln (r/r_s)$. Such terms are associated to the renormalization group running of the Love numbers and are indeed generic in $D>4$ or for dynamical responses. Given that such terms do not change the transformation rules of the response under the spurionic symmetry, our argument shows that they must be absent, hence providing the non-renormalization of the static linear tides. As a matter of fact, this is a requirement to even make sense of the statement that the Love numbers vanish.
We will later present an alternative argument for the non-renormalization using the point-particle EFT, which captures universal effects, such as this running.

The argument easily generalizes to nonlinear  perturbations.
Consider the nonlinear static tidal response to external sources as at $n$-th order in the $w$ sources, which schematically takes the form
\begin{equation}
w_{\rm tail} \sim \,  w_{\rm src}^n 
\left(\dots + c_m \, X^{m} + \dots  \right),
\end{equation}
where $c_m$ is a numerical constant. Under the transformation $w\to -w, \; X \to -X$, the left-hand side is odd, while the term $X^m$ on the right-hand side picks up a factor $(-1)^{n+m}$. Thus we find that the symmetry implies that for even (odd) $n$ all the even (odd) powers of $X$ in the response must vanish.  We can also consider the response of the spatial metric $\gamma_{ij}$. Even if not independent from the $w$ response in $D=4$, this is convenient since it allows us to isolate terms with a different scaling in $X$ and separate the tidal response induced by Love numbers from the gravitational nonlinearities. The $\gamma_{ij}$ tail in the presence of $n$ sources $w_{\rm src}$ is of the form
\begin{equation}
\gamma_{\rm tail} \sim  \,  w_{\rm src}^n 
\left(\dots + g_m \, X^{m} + \dots  \right) \, .
\end{equation}
By the same argument, for even (odd) $n$ all the odd (even) powers of $X$ in the response must vanish. Similar selection rules follow for terms in the response with powers of $\gamma_{\rm src}$. We will explain below that the nonlinear Love numbers correspond to particular terms in the $X$ expansion of the non-linear responses which are set to zero by these selection rules. Therefore, this simple argument will imply vanishing (and non-renormalization) of all non-linear tides for Schwarzschild black holes.

In our argument above we have assumed that the response is an analytic function of $X$ in the far zone. In $D=4$ dimensions the expansion in $X$ is equivalent to expanding in $GM/r$, and its validity is equivalent to the well-known existence of the post-Minkowskian (or Post-Newtonian) expansion in GR. As we shall see later, the situation is more subtle in higher dimensions, where the response is analytic in $1/r$ but can have a non-analytic dependence on $X$. In that case one needs to distinguish between non-renormalization and vanishing of the tidal responses.

We end this section by noting that it is not obvious that the definition of the Love numbers we have used so far is unambiguous or even gauge invariant, especially beyond the linear level. We have used a particular parameterization of the gravitational field and specific coordinates to carry out our arguments. The point-particle EFT described in the next section, provides an unambiguous gauge-invariant definition of the Love numbers and it will allow us to interpret the above zeros. It is therefore important (one might argue that it is even required) to understand how the symmetries used in this section act in the EFT and check consistency with the arguments in this section.

{\bf \em Worldline EFT and tidal operators.--} Tidal responses are usefully described using the language of EFT. At distances $r\gg R$, where $R$ is the size of a black hole or compact gravitating object, the appropriate effective description is one in which such system is replaced by a point particle with action,
\begin{equation}
\label{eq:Spp}
S_{\rm pp} = - M \int d\tau \sqrt{-g_{\mu\nu}\dot x^\mu \dot x^\nu}\, ,
\end{equation}
where $\tau$ is an affine parameter for the worldline (not to be confused with proper time), and $\dot x^\mu = dx^\mu/d\tau$.
This action should be understood as the leading term in a effective theory~\cite{Goldberger:2004jt}, and so it must be supplemented by all higher-dimension operators allowed by the symmetries, which are suppressed by powers of $R$ and encode finite-size effects. The leading such corrections would be couplings of the curvature to intrinsic multipolar gravitational moments on the worldline, which by dimensional analysis should be of size $Q_\ell \sim M R^\ell$. The absence of such multipoles is what distinguishes a black hole from a generic compact object, so we will set them to zero henceforth.

The next allowed operators correspond to higher powers of the curvature, which encode tidal effects. Indeed, in the EFT the tidal response coefficients can be defined in a gauge-invariant way as the Wilson coefficients of such higher-dimension operators. Focusing on static tides, the linear response is characterized by the operators
\begin{align}
S_{\rm Love}^{(1)} = \int d\tau \sum_\ell \Big[
&\lambda^{(E)}_\ell (\partial E)_\ell^2 +\lambda^{(B)}_\ell (\partial B)_\ell^2
  \Big],
\end{align}
Here we use a shorthand notation 
\begin{equation}
(\partial E)_\ell \equiv\partial_{\langle i_1}\dots \partial_{i_{\ell-2}}E_{i_{\ell-1}i_\ell\rangle} \; ,
\end{equation}
and similarly for $(\partial B)_\ell $, where $\langle \cdots \rangle$ denotes the symmetric traceless combination of indices,  and $E_{ij}$ and $B_{ij}$ are the electric and magnetic parts of the Weyl tensor, 
\begin{equation}\label{eq:Weyl}
    E_{\mu \nu} = C_{\mu\rho\nu\sigma} \dot x^\rho \dot x^\sigma\,,\quad B_{\mu \nu} = \tilde{C}_{\mu\rho\nu\sigma} \dot x^\rho \dot x^\sigma\,.
\end{equation}
The mixed $EB$ Love numbers vanish because of parity.

Similarly, we can consider nonlinear static Love numbers.  
The nonlinear static Love numbers, at $n$-th order, are defined as
\begin{align}\label{eq:nonlinearLove}
S_{\rm Love}^{(n)} = \int d\tau \sum_{\vec \ell} \Bigg[
&\lambda^{(E)}_{\vec{\ell},n} \Big( (\partial E)_{\ell_1} \dots (\partial E)_{\ell_{n+1}}  \Big) +\nonumber\\
&\lambda^{(B)}_{\vec{\ell},n} \Big( (\partial B)_{\ell_1} \dots (\partial B)_{\ell_{n+1}}  \Big) +\\
 \sum_{k} \lambda^{(EB)}_{\vec{\ell},n,k} \Big( (\partial B)_{\ell_1} \dots &(\partial B)_{\ell_k} (\partial E)_{\ell_{k+1}} \dots (\partial E)_{\ell_{n+1}} \Big) 
  \Bigg],\nonumber
\end{align}
where we denote with $\vec{\ell}$ an ${(n+1)}$-dimensional multi-index of angular momenta ${\vec{\ell}\equiv (\ell_1, \dots, \ell_{n+1})}$,  and $k$ is an integer $1\leq k\leq n$. Parity invariance restricts $k$ to be even. However our argument will not rely on this, so we keep $k$ arbitrary.  
We are using a compact notation, where we do not write explicitly the index contractions. For any given order in fields and derivatives, the number of independent operators is always finite and can be determined with Hilbert series techniques~\cite{Benvenuti:2006qr,Henning:2017fpj,Haddad:2020que,Aoude:2020ygw}.
For $n=1$, this operator basis reduces the standard linear-response basis with $\vec{\ell}=(\ell,\ell)$. Notice also that since in Eq.~\eqref{eq:nonlinearLove} the indices are contracted with a two-index tensor (the spatial metric $\gamma_{ij}$), a contraction is possible only if $\vert \vec \ell \vert = \sum_{i=0}^{n+1} \ell_i$ is even. Note that all Love numbers with an odd number of $B$ vanish due to parity~\cite{Hadad:2024lsf}.

Power counting (i.e., dimensional analysis) dictates the following scaling of the tidal coefficients 
\begin{equation}
\label{eq:lambdascaling}
    \lambda_{\vec \ell,n} \sim M R^{\vert \vec \ell \vert}\,.
\end{equation}
For a black hole $R=R_S$, so the vanishing of any of them appears to violate naturalness~\cite{Porto:2016zng}. 

So far we have only reviewed the worldline EFT, so let us now understand how the relevant symmetries feature in this description. Let us choose for simplicity a reference frame in which the black hole is at rest, with velocity $\dot x^\mu = \delta^{\mu}_0$, in which
\begin{equation}\label{eq:pp}
S_{\rm pp} = - M \int d\tau e^\phi = - M \int d\tau \,(1+\phi + \dfrac{\phi^2}{2}+\dots)  \;.
\end{equation}
The linear coupling of $\phi$ to the worldline is the only coupling needed to reconstruct the large $r$ expansion of the nonlinear background solution in Eq.~\eqref{eq:wgSchw} using the EFT~\cite{Kol:2011vg,Ivanov:2022hlo}. In terms of our complex field $w$ we can rewrite the point-particle coupling as
\begin{equation}
\label{eq:ppM}
S_{\rm pp} = - M  \int d\tau \,{\rm Re}\, w + \dots \; .
\end{equation}
This coupling breaks explicitly the ${\rm SL}(2,\mathbb{R})$ symmetry of the theory. In particular, it breaks the $\mathbb{Z}_2^S$ symmetry $w\to - w$ unless we assign a spurionic transformation to the mass $M\to -M$, consistently with our previous discussion.\footnote{The worldline terms in $S_{\rm pp}$ that are nonlinear in the scalar are irrelevant for the response computation, as they only affect the matching of the sources and the field variable between full GR and the EFT. We will explain this in detail in~\cite{LongTidesPaper}.} Parity is left unbroken.

We can now derive selection rules on Love numbers by looking at the transformation law of operators under the spurionic symmetry $w \to -w$, $M\to -M$. 
To this end we can express the curvatures in terms of $w$ and $\gamma$. At leading order:
\begin{align}\label{eq:EBw}
&E_{ij} -i B_{ij} = \partial_i \partial_j w + (\partial^2 \gamma)_{ij} \dots \,,\\
&E_{ij}  + i B_{ij} = \partial_i \partial_j \bar w +(\partial^2 \gamma)_{ij} + \dots \,.
\end{align}
See Appendix~\ref{app:dualization} for complete expressions. In the case of operators containing one or more electric tensor $E_{ij}$, we can immediately notice that they do not have homogeneous transformation rules under the symmetry, since $E_{ij}$ depends linearly on both $w$ and $\gamma$. Therefore, we can always rule out this operator if it is generated by dynamics that respects the spurionic symmetry, as is the case for Schwarzschild black holes.

For purely magnetic operators, instead, we start by noticing that parity constrains $n$ to be odd. Moreover, the coefficient $\lambda$ scales always as $M R^{\vert \vec \ell \vert }$, and in the case of black holes in $D=4$ this corresponds always to an odd power of $M$. Putting everything together, under our spurionic symmetry the operator transforms as ${(-)^{n+1}=(+)}$, but the Wilson coefficients pick up a $(-)$ factor. This forces the coefficient to vanish.

We can summarize these findings by saying that in the case of Schwarzschild black holes in four dimensions, all the Love numbers operators break the spurionic $\mathbb{Z}_2^S$ symmetry of the static sector of GR, and are thus forbidden.

We notice here that previous arguments~\cite{Kehagias:2024rtz,Combaluzier-Szteinsznaider:2024sgb} have constrained an infinite family of nonlinear Love numbers, i.e. those probed by axially symmetric perturbations. However this left unconstrained all operators involving $B$, since $B_{ij}=0$ in the presence of axisymmetry. Moreover, the electric Weyl tensor of axisymmetric perturbations satisfies some special relations that leave an infinite number of purely electric Love number operators unconstrained. For instance, $({\rm Tr} E^2)^3 - 6 ({\rm Tr} E^3)^2  =0$ with axisymmetry~\cite{Combaluzier-Szteinsznaider:2024sgb}. Our results go beyond this and apply to general nonlinear static tides involving $E$ and~$B$.

{\bf \em EFT matching and non-renormalization in detail.--} Let us now explain the correspondence between our argument for the selection rules in the full theory and in the Effective Field Theory. The calculation of the Wilson coefficients $\lambda$ which correspond to the Love numbers is carried out by matching a calculation of the tidal response for a black holes in full GR and, perturbatively, in the worldline EFT. That is, by requiring that the responses in GR and the EFT agree
\begin{equation}
    w_{\rm tail}^{\rm GR} = w_{\rm tail}^{\rm EFT}(\lambda)\,, \quad   \gamma_{ij,{\rm tail}}^{\rm GR} = \gamma_{ij,{\rm tail}}^{\rm EFT}(\lambda)\,.
\end{equation}
The latter are a function of the tidal coefficients, $\lambda$, which are fixed by requiring the equality.\footnote{An alternative manifestly gauge-invariant matching can be performed using scattering amplitudes, as was recently explored in~\cite{Ivanov:2024sds,Caron-Huot:2025tlq}.} Such matching allows for an unambiguous determination of the tidal coefficients, even when there is apparent mixing between sources and responses, dissipative and conservative effects, or the true tides and non-linearities (see below and e.g.,~\cite{Rai:2024lho,Ma:2024few}).

The calculation of the responses in full GR is done by solving the equations of motion for perturbations around a Schwarzschild background in the presence of a tidal source field and with appropriate boundary conditions at the horizon, as described in a previous section. 

In the EFT, calculations are done perturbatively in $X=R_S/r$ and the worldline couplings effectively encode the boundary conditions at $r=0$. This perturbative expansion can be conveniently described in the language of Feynman diagrams. 
The calculation of the responses is carried out by turning on sources, $w_\alpha\sim r^\alpha$, and computing diagrams order by order in the source. By appropriately choosing the source terms $w_\alpha$ (their real and imaginary parts, and their tensor structure) we can probe all the nonlinear responses corresponding to different vectors $\vec \ell$. For instance, the $n$-th order tidal responses to external $w$ sources, take the form
\begin{align}
w^{\rm EFT}_{\rm tail} &\sim \, \left( \prod_{\alpha=1}^n w_{\alpha} \right)
\left(
\dots + k^w_{\vec \ell,n} \, X^{\vert \vec \ell \vert +1} + \dots  \right),\\
\gamma^{\rm EFT}_{\rm tail} &\sim \, \left( \prod_{\alpha=1}^n w_{\alpha} \right)\,
\left(
\dots + k^\gamma_{\vec \ell,n} \, X^{\vert \vec \ell \vert +1} + \dots  \right) \, .
\end{align}
where we have singled out the relevant terms, and the remaining powers of $X$ are in the $\cdots$.
Noting that, as explained above, $|\vec\ell|+1$ is an odd number due to rotational invariance, the spurion argument applied to the responses implies that
\begin{align}
    k_{\vec \ell, n}^w &= 0 \quad \text{for } n \text{ odd}\,,\\
    k_{\vec \ell, n}^\gamma &= 0\quad \text{for } n \text{ even}\,.  
\end{align}

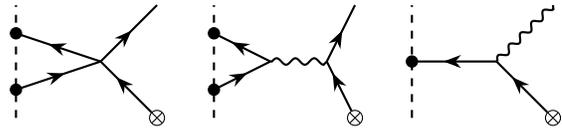
\begin{figure}
    \centering
    \begin{tikzpicture}[baseline={([yshift=-0.4 ex]current bounding box.center)}, scale=0.75]
        \begin{feynman}
            \vertex (i) at (0,0);
            \vertex (e) at (0,2);
            \vertex[dot] (f3) at (0,0.5) {};
            \vertex[dot] (f1) at (0,1.5) {};
            \vertex (c1) at (1.5,1.0);
            \vertex (c3) at (2.5,2.0);
            \vertex[source] (c4) at (2.5,0.0) {};

            \diagram*{
                (i) -- [wl] (e),
                (f3) -- [w] (c1),
                (f1) -- [wbar] (c1),
                (c1) -- [w] (c3),
                (c1) -- [wbar] (c4)
            };
        \end{feynman}
    \end{tikzpicture}
    \hspace{10pt}
    \begin{tikzpicture}[baseline={([yshift=-0.4 ex]current bounding box.center)}, scale=0.75]
        \begin{feynman}
            \vertex (i) at (0,0);
            \vertex (e) at (0,2);
            \vertex[dot] (f3) at (0,0.5) {};
            \vertex[dot] (f1) at (0,1.5) {};
            \vertex (c1) at (1,1.0);
            \vertex (c2) at (2,1.0);
            \vertex (c3) at (2.5,2.0);
            \vertex[source] (c4) at (2.5,0.0) {};

            \diagram*{
                (i) -- [wl] (e),
                (f3) -- [w] (c1),
                (f1) -- [wbar] (c1),
                (c1) -- [sigma] (c2),
                (c2) -- [w] (c3),
                (c2) -- [wbar] (c4)
            };
        \end{feynman}
    \end{tikzpicture}
    \hspace{10pt}
     \begin{tikzpicture}[baseline={([yshift=-0.4 ex]current bounding box.center)},scale=0.75]
        \begin{feynman}
            \vertex (i) at (0,0);
            \vertex (e) at (0,2);
            \vertex[dot] (f1) at (0,1.0) {};
            \vertex (c1) at (1.5,1.0);
            \vertex (c3) at (2.5,2.0);
            \vertex[source] (c4) at (2.5,0.0) {};

            \diagram{
                (i) -- [wl] (e),
                (f1) -- [wbar] (c1),
                (c1) -- [sigma] (c3),
                (c1) -- [wbar] (c4)
            };
        \end{feynman}
    \end{tikzpicture}
    \caption{Diagrams which compute the contribution from Einstein-Hilbert to the linear responses to a source $\bar w$ (denoted by $\otimes$). The two diagrams on the left compute the leading nonlinearity in the response of $w$ (denoted by a solid directed arrow), and the diagram on the right computes the $\gamma_{ij}$ (denoted by a wiggly line) response. }
    \label{fig:responseEH}
\end{figure}

Let us now interpret this as it relates to the Love number operators in the EFT. Straightforward power-counting reveals that these operators contribute to the tidal response at order $|\vec\ell|+1$ in the small $X$ expansion, and hence the terms $k_{\vec \ell,n}$ depend linearly on the Wilson coefficients $\lambda_{\vec\ell,n}$. However, $|\vec\ell|+1$ insertions of the point-particle coupling might also contribute to the tidal response at the same order. To perform a consistent matching it is therefore a priori necessary to compute these contributions, that are formally equivalent to a set of $|\vec\ell|$-loop integrals. This has been studied explicitly in some examples for the case of a minimally coupled scalar and vector in a Schwarzschild background, where it has been shown that individual diagrams are non-vanishing but their sum gives zero to the lowest orders, implying the absence of running for the leading linear tides~\cite{Kol:2011vg,Ivanov:2022hlo}.  In the gravitational case, this has been assumed to be the case, but no symmetry explanation for the non-renormalization or explicit calculation of the nonlinearities (to $|\vec\ell|$-loop order) is currently available. 

Our analysis closes this loophole, as the static symmetry of the EH action, implies that the nonlinearities from diagrams such as those in Fig.~\ref{fig:responseEH} can only affect the non-linear response of $w$ at odd orders in $X$ for even $n$, and that of $\gamma$ at odd $n$.  Crucially, the same tidal coefficient encodes both scalar and tensor responses, due to gauge invariance, but only one of them mixes with the nonlinearities sourced by the coupling in Eq.~\eqref{eq:ppM}.
For instance, the leading non-linear ($n=2$) purely electric response contains terms schematically of the form
\begin{equation}
    (E)^3\supset (\partial^2 w)^3 + (\partial^2 w)^2 (\partial^2 \gamma) + \cdots\,,
\end{equation}
with the same coefficient $\lambda^{(E)}_{(2,2,2),2}$
so it contributes to both $k^{w}_{(2,2,2),2}$ and $k^\gamma_{(2,2,2),2}$. The contribution of the tidal operator to the former mixes with the nonlinearities from the point-mass coupling, but does not for the latter, so we find that
\begin{equation}
    k^\gamma_{(2,2,2),2} \propto  \lambda^{(E)}_{(2,2,2),2} = 0\,.
\end{equation}

Indeed, this argument generalizes, as for even $n$, a tensor response can always be generated in the presence of a non-vanishing Love number, and for odd $n$, a scalar response would be generated.\footnote{Purely magnetic operators exist only for odd $n$, by parity, while electric or mixed type operators include tensor perturbation, see eq.~\eqref{eq:Eij}}  This confirms that our argument \emph{proves vanishing (and non-renormalization) of all non-linear tides for Schwarzschild black holes.}

More generally, our EFT argument implies that non-linearities from the leading point-particle coupling (i.e. ``loops''), which are universal, do not induce any running even when $k_{\vec \ell,n}$ has a non-vanishing matching contribution, as it happens for more general compact objects such as neutron stars. That is, that Love numbers for generic spherically symmetric compact objects are not (logarithmically) renormalized.\footnote{They might, however, receive scheme-dependent power corrections if one uses a scheme that breaks the $\mathbb{Z}_2$ symmetry.}

{\bf \em The higher-dimensional case.--} We can apply similar ideas to study the static response of spherical symmetric black holes in higher-dimensional gravity. We start from the KK parameterization (see Appendix~\ref{app:higherD}):
\begin{equation}\label{eq:temporalKKD}
ds^2 = - e^{2\phi} \left(dt - A_i dx^i \right)^2 + e^{-\frac{2\phi}{D-3}} \gamma_{ij} dx^i dx^j,
\end{equation}
in which the Einstein-Hilbert action (ignoring time derivatives) takes the form 
\begin{align}
S_{\rm EH} =& - \dfrac{1}{16\pi G} \int dt d^{D-1}{x} \sqrt{\gamma} \; \Bigg(-R^{(D-1)}[\gamma] \,+ \\
&\phantom{+}\left(\tfrac{D-2}{D-3}\right) \gamma^{ij} \partial_i\phi \partial_j\phi - \dfrac{1}{4}e^{2\frac{D-2}{D-3}\phi} \gamma^{ik}\gamma^{jl} F_{ij} F_{kl}\Bigg), \nonumber
\end{align}
while the worldline coupling~\eqref{eq:pp} is unchanged. Regarding finite size effect, in higher dimensions the Weyl tensor is decomposed in electric $E_{ij} = C_{0i0j}$, magnetic $B_{ijk} = C_{0ijk}$, and tensor-type operators, with the latter corresponding to a purely spatial Weyl tensor $C_{ijkl}$.\footnote{In $D=4$ these components are not independent and are fixed in terms of $E_{ij}$ and $B_{ij}$. This is no longer the case for $D>4$.} There are, therefore, additional static Love-number operators built out of
\begin{equation}
(\partial C)_\ell \equiv\partial_{\langle i_1}\dots \partial_{i_{\ell-2}}C_{i_{\ell-1}i_\ell\rangle jk} \; ,
\end{equation}
whose coefficients we denote as $\lambda^{(T)}$ for purely tensor, and $\lambda^{(TE)}$, $\lambda^{(TB)}$ for mixed response.

At leading order, the electric and tensor-type operators are associated to $\phi$ and $\gamma_{ij}$ respectively, while the magnetic operator is associated to $A_i$. Moreover, since due to parity $A_{i}$ (or $F_{ij}$) enters quadratically in the action, and the point-particle worldline coupling only involves $\phi$, it is easy to see that the $F^2$ term in the action does not contribute at the classical level to all orders in the $G$ expansion, unless $A_i$ is attached to an external leg (source or tail). That is: the operator $F^2$ can be neglected when considering the electric or tensor-type static response.  We leave an analysis of the magnetic response of higher dimensional black holes to future work. 

In this case, the transformation $\phi \to -\phi$ ($\gamma_{ij}$ invariant) is a symmetry of the action, and remains a symmetry of the background solution, the Schwarzschild--Tangherlini black hole, if we transform $M\to -M$, see Appendix~\ref{app:higherD}. We denote this symmetry as $\mathbb{Z}_2^{\phi}$. We can therefore use arguments similar to the ones we presented before to constrain the electric and tensor-type static Love numbers in higher dimensions. Dimensional analysis dictates that the response admits an expansion in the variable (see Appendix~\ref{app:higherD}):
\begin{equation}\label{eq:X_higherD}
X= \dfrac{1}{4}\left(\dfrac{R_S}{r}\right)^{D-3},
\end{equation} 
where in particular $X\propto GM$.

A straightforward power counting argument reveals that the linear Love numbers $\lambda_\ell$ contribute to the tidal response at order $1/r^{\ell+D-3}$. In the presence of a source, $r^\ell$, and in terms of the dimensionless Love numbers $k_\ell$:
\begin{equation}
g_{\rm tail} \sim \,\bar{g} \left(1+\dots + k_\ell \, X^{2 \hat{\ell}+1} + \dots  \right) ,
\end{equation}
where $\bar{g} \sim r^\ell$ is the gravitational tidal source, and ${\hat{\ell}= \ell/(D-3)}$. Such a response function can be generated perturbatively in $G$ only if $2\hat{\ell} \in \mathbb{N}$. The $\mathbb{Z}_2$ symmetry $\phi\to-\phi$, $M\to-M$ implies then the vanishing of Love numbers for integer $\hat\ell$, while allowing a running for $\hat\ell = (2N+1)/2$, with $N\in \mathbb{N}$. This matches exactly the pattern found in~\cite{Kol:2011vg,Hui:2020xxx,Hadad:2024lsf}.

The above argument also implies that Love numbers with $2\hat\ell \notin \mathbb{N}$ do not run. Their matching, on the other hand, is associated to non-perturbative (i.e. non-analytic) contributions in $GM$. Indeed, the scaling of the tidal coefficients with $R_S$ is dictated by dimensional analysis, and due to eq.~\eqref{eq:X_higherD} this corresponds to an integer power of $GM$ only when $2\hat\ell$ is integer.  This is in contrast to $D=4$ dimensions, where all terms in the response are perturbative in $G$. In general $D$, the nonperturbative contributions are determined directly in the matching computation and are unconstrained by our spurion analysis. Our arguments imply only their non-renormalization. This is in complete analogy with perturbative non-renormalization theorems in supersymmetric gauge theories~\cite{Affleck:1983mk}.

We can generalize the same analysis to the nonlinear static response of electric and tensor type. In this case power counting dictates that the static Love number $\lambda_{\vec \ell,n}$ contributes to the tidal response at order $1/r^{\ell_{\rm tail} +(D-3)}$. In terms of dimensionless Love numbers this gives \begin{equation}
g_{\rm tail} \sim \, \left( \prod_{\alpha=1}^n g_{\alpha} \right)
\left(c_n+\dots + k_{\vec \ell,n} \, X^{\vert \vec \ell \vert /(D-3) +1} + \dots  \right),
\end{equation}
where $\vert \vec \ell \vert = \sum_{i=1}^{n+1} \ell_i$ and $ g_{\alpha} \sim r^{\ell_\alpha}$ are the $n$ gravitational tidal sources. 
This response can be generated perturbatively only when $\vert \vec \ell \vert \in (D-3)\mathbb{N}$.
The selection rules following from the $\mathbb{Z}_2^{\phi}$ symmetry are easily determined depending on the value of $\vert \vec \ell \vert$, and on whether $g_{\rm tail}$ and the $g_\alpha$'s are $\phi$'s or $\gamma$'s. 

For example, for purely electric Love numbers: Love numbers do not run, and vanish in the case of a $\mathbb{Z}_2^{\phi}$ symmetric background such as the Schwarzschild--Tangherlini solution, whenever ${\vert \vec \ell \vert/(D-3)} \equiv n+1$ (mod~$2$), while a running is allowed for ${\vert \vec \ell \vert/(D-3)} \equiv n$ (mod~$2$).
For purely tensor Love numbers: $k_{\vec\ell ,n}$  can run only when $\vert \vec \ell \vert/({D-3})$ is an odd integer.

{\bf \em A generalization for shift symmetric scalars: no-hair and no-tides.--} We can also apply our arguments to more general theories involving additional fields. As a simple example, we consider here the case of a shift symmetric scalar field $\psi$ in a Schwarzschild background. 
We restrict to leading order in derivatives and minimal coupling for simplicity (the extension to Horndeski-type operators~\cite{Horndeski:1974wa,Deffayet:2011gz} is straightforward). In this case, the action is just 
\begin{equation}\label{Spsi}
S_\psi= - \int d^4x\sqrt{\gamma} \,e^{-2\phi}\,P\Big(e^{2\phi}\gamma^{ij}\partial_i \psi \partial_j \psi \Big),
\end{equation} 
where $P(Y)$ is an arbitrary function which admits a Taylor expansion around $0$ which starts as $Y/2$, giving $\sqrt{\gamma} (\partial_i \psi)^2/2$.
The shift symmetry sets to zero the minimal coupling of the massive point-particle to $\psi$, as well as the linear couplings $\int d\tau (\square)^n\psi$ which vanish on-shell in the gaussian theory and can be removed by a field redefinition. On the other hand, the nonlinear tidal response is defined by
\begin{equation}
S_{\rm \psi,Love}^{(n)} = \int d\tau \sum_{\vec \ell} 
\lambda^{(\psi)}_{\vec{\ell},n} \Big( (\partial \psi)_{\ell_1} \dots (\partial \psi)_{\ell_{n+1}}  \Big) ,
\end{equation}
where 
\begin{equation}
(\partial \psi)_\ell \equiv\partial_{\langle i_1}\dots \partial_{i_{\ell}\rangle}\psi \,.
\end{equation}
Also in this case, $\lambda_{\vec \ell, n}=0$ for $\vert\vec\ell\vert$ odd.
In the case of a free scalar, $P(X)=X/2$, the scalar $\psi$ interacts directly only with $\gamma_{ij}$ and its action is invariant under the ${\rm SL}(2,\mathbb{R})$ symmetry of the gravitational action.

A straightforward power counting reveals that the static Love number $\lambda^{(\psi)}_{\vec \ell,n}$ contributes to the tidal response at order $1/r^{\ell_{\rm tail} +n}$:
\begin{equation}
\psi_{\rm tail} \sim \,  \left( \prod_{\alpha=1}^n \psi_{\alpha} \right) \left(c_n+\dots + k^{(\psi)}_{\vec \ell,n} \, X^{\vert \vec \ell \vert +1} + \dots  \right),
\end{equation}
where $\vert \vec \ell \vert = \sum_{i=1}^{n+1} \ell_i$, and $ w_{\alpha} \sim r^{\ell_\alpha}$ ($\alpha=1,\dots n$) are the $n$ gravitational tidal sources. 
For simplicity of notation we suppressed the tensor structures, but those are in one-to-one correspondence with the possible contractions  of order $n$ (equivalently, with independent operators up to equations of motion and integration-by-parts identities).

When $n$ is even $\psi_{\rm tail}$ is forced to vanish by invariance under $\psi\to -\psi$, which is an exact $\mathbb{Z}_2^{\psi}$ symmetry of the action~\eqref{Spsi} and is unbroken by the point-particle worldline coupling. This is the case also if there is \emph{no source} ($n=0$), in which case our statement reduces to the \emph{no-hair theorem} for shift symmetric scalar fields~\cite{Hui:2012qt}.
Here we are assuming that $\psi \to 0$ at infinity, \emph{i.e.} the $\mathbb{Z}_2^{\psi}$ symmetry is not spontaneously broken asymptotically, and regularity at the horizon. 

For $n$ odd, instead, the tidal response vanishes only for the case of a free scalar, or for $n=1$. Indeed, the action~\eqref{Spsi} breaks the gravitational symmetry $\mathbb{Z}_2^{S}$, $w\to -w$, unless the action of $\psi$ is quadratic. Therefore, it only has vanishing tidal response for a probe minimally-coupled scalar thanks to the symmetry ${w\to -w, \; M \to -M}$, which is an exact symmetry of the gravitational action and of the background solution. We thus recover the vanishing result for the linear response of a probe minimally-coupled scalar~\cite{Kol:2011vg,Ivanov:2022hlo,santoni2024black}. 
More generally, the linear static Love number ($n=1$) is always vanishing even in the presence of shift-symmetric self-interactions, since only the quadratic term in the action contributes classically and this is invariant under $\mathbb{Z}_2^{S}$.
On the other hand, self-interactions break the symmetry so that the nonlinear response can be non-vanishing for all odd $n\geq 3$, in agreement with the explicit result of~\cite{DeLuca:2023mio} for $n=3$. Notice also that the addition of a mass term for the scalar breaks explicitly the gravitational $\mathbb{Z}_2^{S}$, and allows a non-vanishing tidal response, consistent with explicit computations~\cite{Santos}. The $\mathbb{Z}_2^{\psi}$, on the other hand, is unbroken and the no-hair property is preserved \cite{Bekenstein:1972ny}.

To summarize, these observations generalize the vanishing results that appeared for the linear response of a probe free scalar in a Schwarzschild background to minimally coupled shift-symmetric scalar fields, and to the nonlinear static tidal response with even $n$. It also unifies the well-known no-hair theorem with the vanishing of Love numbers. 

The assumptions in our no-hair statement match those of~\cite{Hui:2012qt}: spherical symmetry, time-independence, asymptotic flatness, and regularity at the horizon. 
We expect these results to admit exceptions as soon as we include non-minimal couplings that break the  symmetries previously discussed. For instance, a coupling of the scalar to the Gauss-Bonnet density, while being shift symmetric on manifolds with vanishing Gauss-Bonnet invariant, explicitly breaks the $\mathbb{Z}_2^{\psi}$ symmetry $\psi \to -\psi$. This provides a symmetry understanding of why hairy solutions are allowed in the presence of such a coupling~\cite{Sotiriou:2013qea,Babichev:2017guv,Creminelli:2020lxn,Hui:2021cpm}.\footnote{Notice, however, that not all $\mathbb{Z}_2^{\psi}$ breaking couplings are guaranteed to generate a regular hairy black hole solution~\cite{Hui:2012qt,Babichev:2017guv,Creminelli:2020lxn}.}

{\bf \em Further extensions.--}
The approach we presented for the study of time-independent perturbations of spherically symmetric black holes can be readily extended to include a gravito-magnetic (NUT) charge and to treat charged black holes in Einstein--Maxwell theory, as we will detail in a forthcoming publication~\cite{LongTidesPaper}. Our arguments are also fully consistent with the finding of non-vanishing fermionic responses of neutral black holes~\cite{Chakraborty:2025zyb} and charged responses of electromagnetically charged black holes~\cite{Ma:2024few,Pereniguez:2025jxq}. In both of these cases, we find that the corresponding accidental symmetries are explicitly broken at the level of the action~\cite{LongTidesPaper}. Similarly, a cosmological constant breaks the symmetry, consistent with the observation that large black holes in (Anti-)de-Sitter space can have non-zero static Love numbers~\cite{Costa:2015gol,Nair:2024mya,Franzin:2024cah}, and hence polarize. (Small black holes have vanishing tides~\cite{Hui:2020xxx}, as expected from the flat-space limit.)
 It is also easy to see that the symmetry will be generically broken in EFT extensions of gravity involving higher dimensional operators, in agreement with~\cite{Cardoso:2018ptl,DeLuca:2024nih,DeLuca:2022tkm,Barbosa:2025uau,Barura:2024uog,Bhattacharyya:2025slf}. Note, however, that in all of these cases the EFT matching computation cannot be performed at tree level, but requires computing nonlinearities in the EFT, so the results in the literature remain incomplete.

{\bf \em Comments on dynamical tides.--} The symmetry we presented is an accidental symmetry of the static sector of GR, and hence it is expected that it will be broken by time-dependent perturbations. This explains why non-vanishing dynamical Love (including dissipation) numbers are allowed, in agreement with explicit computations (see e.g.,~\cite{Goldberger:2019sya,Charalambous:2022rre,Saketh:2023bul}). 
For instance, the leading dynamical tides are dissipative and correspond to EFT operators of the form 
\begin{align} \label{eq:disstides}
S_{\rm dis.} = \int d\tau \sum_\ell \Big[
&\lambda^{(E)}_{\ell,\omega} (\partial \dot E_-)_\ell(\partial E_+)_\ell  
+ \cdots 
  \Big]\,. 
\end{align}
Here we work in the \emph{in-in} formalism with doubled degrees of freedom denoted by subscripts $\pm$ (see e.g.,~\cite{Caron-Huot:2025tlq}), the dot denotes a derivative with respect to $\tau$, and the $\cdots$ include magnetic and mixed, as well as nonlinear, terms.

These responses can be calculated by a matching computations with time dependent sources   with frequency $\omega$, $w\sim e^{-i\tau\omega}$~\cite{Goldberger:2019sya,Charalambous:2022rre}. The frequency introduces an additional scale into the problem, and in particular it allows terms in the response of the form
\begin{equation}
w_{\rm tail} \supset \, \left( \prod_{\alpha=1}^n w_{\alpha} \right) k_{\vec \ell,n,\omega}\, (R_S\omega)\, X^{\vert \vec \ell \vert +1} \,,
\end{equation}
which are now allowed by the spurionic symmetry, due to the additional power of $R_S\sim M$. The same is true for more general dynamical tides, with more time derivatives. In fact, such Love numbers undergo renormalization group running~\cite{Mandal:2023hqa,Jakobsen:2023pvx,Saketh:2023bul,Ivanov:2024sds,Caron-Huot:2025tlq}, which appears as logarithmic terms in the response.  In particular, this means that their value is only meaningful at a given scale and by explicit matching to the EFT~\cite{Goldberger:2019sya,Ivanov:2024sds,Caron-Huot:2025tlq}. 

We note, however, that the breaking of the accidental symmetry is given by terms with even numbers of time derivatives in Eq.~\eqref{eq:zaction} due to time-reversal symmetry of the Einstein-Hilbert action. Thus, the breaking of time reversal in the response can only arise from the boundary conditions at the horizon in full GR, or from the tidal operators in the EFT. As a consequence, the leading dissipative tides in Eq.~\eqref{eq:disstides} and their non-linear analogs are not renormalized. We will explain this in more detail in upcoming work.

{\bf \em Conclusions.--} We presented an accidental symmetry of the static sector of GR and explained how it can be used to prove selection rules on static Love number, valid fully nonlinearly in classical gravity. We used this to show that Schwarzschild black holes have vanishing static nonlinear Love numbers and, more generally, that Love numbers do not run in the worldline EFT for generic compact objects. We also discussed how similar considerations can be used to explain the pattern of vanishing and running Love numbers of electric and tensor type for spherically symmetric black holes in higher-dimensional gravity, generalizing known results to nonlinear level. When applied to the EFT of a shift-symmetric scalar field coupled to gravity, we have shown that our approach provides an alternative symmetry-based proof of the no-hair theorem, unifying it with the vanishing of the static tidal response.  In upcoming work we will detail how our arguments extend to electromagnetically and NUT charged black holes~\cite{LongTidesPaper}, and the generalization to spinning holes is in progress. This gives us confidence that our arguments provide a uniform long-sought solution to the naturalness problem of vanishing black-hole tides. 

Looking forward, it would be very interesting to understand how the vanishing of nonlinear tides is imprinted in the gravitational-wave signals from binary black-hole mergers, particularly in the strong-field region near merger, which would provide the most stringent test of this striking prediction of Einstein's theory.

\vspace{5pt}
{\bf \em Acknowledgements.} We would like to thank Donato Bini, Clifford Cheung, Thibault Damour, Lam Hui, Mikhail Ivanov, Austin Joyce, Zohar Komargodski, Yue-Zhou Li, Rafael Porto, Riccardo Rattazzi, Ira Rothstein, Luca Santoni, Jorge Santos and Zihan Zhou for useful discussions. We also thank Tom Hartman, Alberto Nicolis, and Luca Santoni for comments on a preliminary version of this article.
The work of A.P. is supported by the Huawei Young Talents Program at IHES.


\bibliographystyle{utphys}
\bibliography{refs.bib}

\providecommand{\href}[2]{#2}\begingroup\raggedright\begin{thebibliography}{10}

\bibitem{Damour:1982wm}
T.~Damour, ``{Gravitational radiation and the motion of compact bodies},'' In
  {\it Les Houches Summer School on Gravitational Radiation},
  \url{https://www.ihes.fr/~damour/publications/27-83.pdf}, 1982.

\bibitem{Fang:2005qq}
H.~Fang and G.~Lovelace, ``{Tidal coupling of a Schwarzschild black hole and
  circularly orbiting moon},''
  \href{https://dx.doi.org/10.1103/PhysRevD.72.124016}{{\em Phys. Rev. D}
  {\bfseries 72} (2005) 124016},
  \href{https://arxiv.org/abs/gr-qc/0505156}{{\ttfamily arXiv:gr-qc/0505156}}.

\bibitem{Damour:2009vw}
T.~Damour and A.~Nagar, ``{Relativistic tidal properties of neutron stars},''
  \href{https://dx.doi.org/10.1103/PhysRevD.80.084035}{{\em Phys. Rev. D}
  {\bfseries 80} (2009) 084035},
  \href{https://arxiv.org/abs/0906.0096}{{\ttfamily arXiv:0906.0096 [gr-qc]}}.

\bibitem{Binnington:2009bb}
T.~Binnington and E.~Poisson, ``{Relativistic theory of tidal Love numbers},''
  \href{https://dx.doi.org/10.1103/PhysRevD.80.084018}{{\em Phys. Rev. D}
  {\bfseries 80} (2009) 084018},
  \href{https://arxiv.org/abs/0906.1366}{{\ttfamily arXiv:0906.1366 [gr-qc]}}.

\bibitem{Damour:2009va}
T.~Damour and O.~M. Lecian, ``{On the gravitational polarizability of black
  holes},'' \href{https://dx.doi.org/10.1103/PhysRevD.80.044017}{{\em Phys.
  Rev. D} {\bfseries 80} (2009) 044017},
  \href{https://arxiv.org/abs/0906.3003}{{\ttfamily arXiv:0906.3003 [gr-qc]}}.

\bibitem{Kol:2011vg}
B.~Kol and M.~Smolkin, ``{Black hole stereotyping: Induced gravito-static
  polarization},'' \href{https://dx.doi.org/10.1007/JHEP02(2012)010}{{\em JHEP}
  {\bfseries 02} (2012) 010}, \href{https://arxiv.org/abs/1110.3764}{{\ttfamily
  arXiv:1110.3764 [hep-th]}}.

\bibitem{Hui:2020xxx}
L.~Hui, A.~Joyce, R.~Penco, L.~Santoni, and A.~R. Solomon, ``{Static response
  and Love numbers of Schwarzschild black holes},''
  \href{https://dx.doi.org/10.1088/1475-7516/2021/04/052}{{\em JCAP} {\bfseries
  04} (2021) 052}, \href{https://arxiv.org/abs/2010.00593}{{\ttfamily
  arXiv:2010.00593 [hep-th]}}.

\bibitem{Ivanov:2022hlo}
M.~M. Ivanov and Z.~Zhou, ``{Revisiting the matching of black hole tidal
  responses: A systematic study of relativistic and logarithmic corrections},''
  \href{https://dx.doi.org/10.1103/PhysRevD.107.084030}{{\em Phys. Rev. D}
  {\bfseries 107} no.~8, (2023) 084030},
  \href{https://arxiv.org/abs/2208.08459}{{\ttfamily arXiv:2208.08459
  [hep-th]}}.

\bibitem{Hadad:2024lsf}
T.~Hadad, B.~Kol, and M.~Smolkin, ``{Gravito-magnetic polarization of
  Schwarzschild black hole},''
  \href{https://dx.doi.org/10.1007/JHEP06(2024)169}{{\em JHEP} {\bfseries 06}
  (2024) 169}, \href{https://arxiv.org/abs/2402.16172}{{\ttfamily
  arXiv:2402.16172 [hep-th]}}.

\bibitem{Poisson:2020vap}
E.~Poisson, ``{Compact body in a tidal environment: New types of relativistic
  Love numbers, and a post-Newtonian operational definition for tidally induced
  multipole moments},''
  \href{https://dx.doi.org/10.1103/PhysRevD.103.064023}{{\em Phys. Rev. D}
  {\bfseries 103} no.~6, (2021) 064023},
  \href{https://arxiv.org/abs/2012.10184}{{\ttfamily arXiv:2012.10184
  [gr-qc]}}.

\bibitem{Poisson:2021yau}
E.~Poisson, ``{Tidally induced multipole moments of a nonrotating black hole
  vanish to all post-Newtonian orders},''
  \href{https://dx.doi.org/10.1103/PhysRevD.104.104062}{{\em Phys. Rev. D}
  {\bfseries 104} no.~10, (2021) 104062},
  \href{https://arxiv.org/abs/2108.07328}{{\ttfamily arXiv:2108.07328
  [gr-qc]}}.

\bibitem{Riva:2023rcm}
M.~M. Riva, L.~Santoni, N.~Savi\'c, and F.~Vernizzi, ``{Vanishing of nonlinear
  tidal Love numbers of Schwarzschild black holes},''
  \href{https://dx.doi.org/10.1016/j.physletb.2024.138710}{{\em Phys. Lett. B}
  {\bfseries 854} (2024) 138710},
  \href{https://arxiv.org/abs/2312.05065}{{\ttfamily arXiv:2312.05065
  [gr-qc]}}.

\bibitem{Iteanu:2024dvx}
S.~Iteanu, M.~M. Riva, L.~Santoni, N.~Savi\'c, and F.~Vernizzi, ``{Vanishing of
  quadratic Love numbers of Schwarzschild black holes},''
  \href{https://dx.doi.org/10.1007/JHEP02(2025)174}{{\em JHEP} {\bfseries 02}
  (2025) 174}, \href{https://arxiv.org/abs/2410.03542}{{\ttfamily
  arXiv:2410.03542 [gr-qc]}}.

\bibitem{Kehagias:2024rtz}
A.~Kehagias and A.~Riotto, ``{Black holes in a gravitational field: the
  non-linear static love number of Schwarzschild black holes vanishes},''
  \href{https://dx.doi.org/10.1088/1475-7516/2025/05/039}{{\em JCAP} {\bfseries
  05} (2025) 039}, \href{https://arxiv.org/abs/2410.11014}{{\ttfamily
  arXiv:2410.11014 [gr-qc]}}.

\bibitem{Combaluzier-Szteinsznaider:2024sgb}
O.~Combaluzier-Szteinsznaider, L.~Hui, L.~Santoni, A.~R. Solomon, and S.~S.~C.
  Wong, ``{Symmetries of vanishing nonlinear Love numbers of Schwarzschild
  black holes},'' \href{https://dx.doi.org/10.1007/JHEP03(2025)124}{{\em JHEP}
  {\bfseries 03} (2025) 124},
  \href{https://arxiv.org/abs/2410.10952}{{\ttfamily arXiv:2410.10952
  [gr-qc]}}.

\bibitem{Gounis:2024hcm}
L.~R. Gounis, A.~Kehagias, and A.~Riotto, ``{The vanishing of the non-linear
  static love number of Kerr black holes and the role of symmetries},''
  \href{https://dx.doi.org/10.1088/1475-7516/2025/03/002}{{\em JCAP} {\bfseries
  03} (2025) 002}, \href{https://arxiv.org/abs/2412.08249}{{\ttfamily
  arXiv:2412.08249 [gr-qc]}}.

\bibitem{Goldberger:2004jt}
W.~D. Goldberger and I.~Z. Rothstein, ``{An Effective field theory of gravity
  for extended objects},''
  \href{https://dx.doi.org/10.1103/PhysRevD.73.104029}{{\em Phys. Rev. D}
  {\bfseries 73} (2006) 104029},
  \href{https://arxiv.org/abs/hep-th/0409156}{{\ttfamily
  arXiv:hep-th/0409156}}.

\bibitem{Porto:2005ac}
R.~A. Porto, ``{Post-Newtonian corrections to the motion of spinning bodies in
  NRGR},'' \href{https://dx.doi.org/10.1103/PhysRevD.73.104031}{{\em Phys. Rev.
  D} {\bfseries 73} (2006) 104031},
  \href{https://arxiv.org/abs/gr-qc/0511061}{{\ttfamily arXiv:gr-qc/0511061}}.

\bibitem{tHooft:1979rat}
G.~'t~Hooft, ``{Naturalness, chiral symmetry, and spontaneous chiral symmetry
  breaking},'' \href{https://dx.doi.org/10.1007/978-1-4684-7571-5_9}{{\em NATO
  Sci. Ser. B} {\bfseries 59} (1980) 135--157}.

\bibitem{Porto:2016zng}
R.~A. Porto, ``{The Tune of Love and the Nature(ness) of Spacetime},''
  \href{https://dx.doi.org/10.1002/prop.201600064}{{\em Fortsch. Phys.}
  {\bfseries 64} no.~10, (2016) 723--729},
  \href{https://arxiv.org/abs/1606.08895}{{\ttfamily arXiv:1606.08895
  [gr-qc]}}.

\bibitem{Charalambous:2021kcz}
P.~Charalambous, S.~Dubovsky, and M.~M. Ivanov, ``{Hidden Symmetry of Vanishing
  Love Numbers},''
  \href{https://dx.doi.org/10.1103/PhysRevLett.127.101101}{{\em Phys. Rev.
  Lett.} {\bfseries 127} no.~10, (2021) 101101},
  \href{https://arxiv.org/abs/2103.01234}{{\ttfamily arXiv:2103.01234
  [hep-th]}}.

\bibitem{Hui:2021vcv}
L.~Hui, A.~Joyce, R.~Penco, L.~Santoni, and A.~R. Solomon, ``{Ladder symmetries
  of black holes. Implications for love numbers and no-hair theorems},''
  \href{https://dx.doi.org/10.1088/1475-7516/2022/01/032}{{\em JCAP} {\bfseries
  01} no.~01, (2022) 032}, \href{https://arxiv.org/abs/2105.01069}{{\ttfamily
  arXiv:2105.01069 [hep-th]}}.

\bibitem{Charalambous:2022rre}
P.~Charalambous, S.~Dubovsky, and M.~M. Ivanov, ``{Love symmetry},''
  \href{https://dx.doi.org/10.1007/JHEP10(2022)175}{{\em JHEP} {\bfseries 10}
  (2022) 175}, \href{https://arxiv.org/abs/2209.02091}{{\ttfamily
  arXiv:2209.02091 [hep-th]}}.

\bibitem{Hui:2022vbh}
L.~Hui, A.~Joyce, R.~Penco, L.~Santoni, and A.~R. Solomon, ``{Near-zone
  symmetries of Kerr black holes},''
  \href{https://dx.doi.org/10.1007/JHEP09(2022)049}{{\em JHEP} {\bfseries 09}
  (2022) 049}, \href{https://arxiv.org/abs/2203.08832}{{\ttfamily
  arXiv:2203.08832 [hep-th]}}.

\bibitem{Berens:2022ebl}
R.~Berens, L.~Hui, and Z.~Sun, ``{Ladder symmetries of black holes and de
  Sitter space: love numbers and quasinormal modes},''
  \href{https://dx.doi.org/10.1088/1475-7516/2023/06/056}{{\em JCAP} {\bfseries
  06} (2023) 056}, \href{https://arxiv.org/abs/2212.09367}{{\ttfamily
  arXiv:2212.09367 [hep-th]}}.

\bibitem{BenAchour:2022uqo}
J.~Ben~Achour, E.~R. Livine, S.~Mukohyama, and J.-P. Uzan, ``{Hidden symmetry
  of the static response of black holes: applications to Love numbers},''
  \href{https://dx.doi.org/10.1007/JHEP07(2022)112}{{\em JHEP} {\bfseries 07}
  (2022) 112}, \href{https://arxiv.org/abs/2202.12828}{{\ttfamily
  arXiv:2202.12828 [gr-qc]}}.

\bibitem{Sharma:2024hlz}
C.~Sharma, R.~Ghosh, and S.~Sarkar, ``{Exploring ladder symmetry and Love
  numbers for static and rotating black holes},''
  \href{https://dx.doi.org/10.1103/PhysRevD.109.L041505}{{\em Phys. Rev. D}
  {\bfseries 109} no.~4, (2024) L041505},
  \href{https://arxiv.org/abs/2401.00703}{{\ttfamily arXiv:2401.00703
  [gr-qc]}}.

\bibitem{Berens:2025okm}
R.~Berens, L.~Hui, D.~McLoughlin, R.~Penco, and J.~Staunton, ``{Geometric
  Symmetries for the Vanishing of the Black Hole Tidal Love Numbers},''
  \href{https://arxiv.org/abs/2510.18952}{{\ttfamily arXiv:2510.18952
  [hep-th]}}.

\bibitem{Geroch:1970nt}
R.~P. Geroch, ``{A Method for generating solutions of Einstein's equations},''
  \href{https://dx.doi.org/10.1063/1.1665681}{{\em J. Math. Phys.} {\bfseries
  12} (1971) 918--924}.

\bibitem{Buchdahl:1954}
H.~Buchdahl, ``{New Solutions of the Einstein‐Maxwell Equations from Old},''
  \href{https://dx.doi.org/10.1093/qmath/5.1.116}{{\em Quart. J. Math.}
  {\bfseries 5 (1)} (1954) 116--119}.

\bibitem{Ehlers:1959aug}
J.~Ehlers, ``{Transformations of static exterior solutions of Einstein's
  gravitational field equations into different solutions by means of conformal
  mapping},'' {\em Colloq. Int. CNRS} {\bfseries 91} (1962) 275--284.

\bibitem{Ernst:1967wx}
F.~J. Ernst, ``{New formulation of the axially symmetric gravitational field
  problem},'' \href{https://dx.doi.org/10.1103/PhysRev.167.1175}{{\em Phys.
  Rev.} {\bfseries 167} (1968) 1175--1179}.

\bibitem{Stephani:2003tm}
H.~Stephani, D.~Kramer, M.~A.~H. MacCallum, C.~Hoenselaers, and E.~Herlt,
  \href{https://dx.doi.org/10.1017/CBO9780511535185}{{\em {Exact solutions of
  Einstein's field equations}}}.
\newblock Cambridge Monographs on Mathematical Physics. Cambridge Univ. Press,
  Cambridge, 2003.

\bibitem{santoni2024black}
L.~Santoni, ``Black hole perturbation theory,''
  \url{https://courses.ipht.fr/sites/default/files/local-media-files--2024-07-12%2019%3A15/BHPT.pdf},
  2024.

\bibitem{Diedrichs:2025vhv}
R.~F. Diedrichs, S.~Tsujikawa, and K.~Yagi, ``{Tidal Love numbers of neutron
  stars in Horndeski theories},''
  \href{https://dx.doi.org/10.1103/cmb4-chn3}{{\em Phys. Rev. D} {\bfseries
  112} no.~4, (2025) 044023},
  \href{https://arxiv.org/abs/2501.07998}{{\ttfamily arXiv:2501.07998
  [gr-qc]}}.

\bibitem{LongTidesPaper}
J.~Parra-Martinez and A.~Podo. To appear.

\bibitem{Benvenuti:2006qr}
S.~Benvenuti, B.~Feng, A.~Hanany, and Y.-H. He, ``{Counting BPS Operators in
  Gauge Theories: Quivers, Syzygies and Plethystics},''
  \href{https://dx.doi.org/10.1088/1126-6708/2007/11/050}{{\em JHEP} {\bfseries
  11} (2007) 050}, \href{https://arxiv.org/abs/hep-th/0608050}{{\ttfamily
  arXiv:hep-th/0608050}}.

\bibitem{Henning:2017fpj}
B.~Henning, X.~Lu, T.~Melia, and H.~Murayama, ``{Operator bases, $S$-matrices,
  and their partition functions},''
  \href{https://dx.doi.org/10.1007/JHEP10(2017)199}{{\em JHEP} {\bfseries 10}
  (2017) 199}, \href{https://arxiv.org/abs/1706.08520}{{\ttfamily
  arXiv:1706.08520 [hep-th]}}.

\bibitem{Haddad:2020que}
K.~Haddad and A.~Helset, ``{Tidal effects in quantum field theory},''
  \href{https://dx.doi.org/10.1007/JHEP12(2020)024}{{\em JHEP} {\bfseries 12}
  (2020) 024}, \href{https://arxiv.org/abs/2008.04920}{{\ttfamily
  arXiv:2008.04920 [hep-th]}}.

\bibitem{Aoude:2020ygw}
R.~Aoude, K.~Haddad, and A.~Helset, ``{Tidal effects for spinning particles},''
  \href{https://dx.doi.org/10.1007/JHEP03(2021)097}{{\em JHEP} {\bfseries 03}
  (2021) 097}, \href{https://arxiv.org/abs/2012.05256}{{\ttfamily
  arXiv:2012.05256 [hep-th]}}.

\bibitem{Ivanov:2024sds}
M.~M. Ivanov, Y.-Z. Li, J.~Parra-Martinez, and Z.~Zhou, ``{Gravitational Raman
  Scattering in Effective Field Theory: A Scalar Tidal Matching at O(G3)},''
  \href{https://dx.doi.org/10.1103/PhysRevLett.132.131401}{{\em Phys. Rev.
  Lett.} {\bfseries 132} no.~13, (2024) 131401},
  \href{https://arxiv.org/abs/2401.08752}{{\ttfamily arXiv:2401.08752
  [hep-th]}}.

\bibitem{Caron-Huot:2025tlq}
S.~Caron-Huot, M.~Correia, G.~Isabella, and M.~Solon, ``{Gravitational Wave
  Scattering via the Born Series: Scalar Tidal Matching to $\mathcal{O}(G^7)$
  and Beyond},'' \href{https://arxiv.org/abs/2503.13593}{{\ttfamily
  arXiv:2503.13593 [hep-th]}}.

\bibitem{Rai:2024lho}
M.~Rai and L.~Santoni, ``{Ladder symmetries and Love numbers of
  Reissner-Nordstr{\"o}m black holes},''
  \href{https://dx.doi.org/10.1007/JHEP07(2024)098}{{\em JHEP} {\bfseries 07}
  (2024) 098}, \href{https://arxiv.org/abs/2404.06544}{{\ttfamily
  arXiv:2404.06544 [gr-qc]}}.

\bibitem{Ma:2024few}
L.~Ma, Z.-H. Wu, Y.~Pang, and H.~Lu, ``{Charging the Love numbers: Charged
  scalar response coefficients of Kerr-Newman black holes},''
  \href{https://dx.doi.org/10.1103/PhysRevD.111.044003}{{\em Phys. Rev. D}
  {\bfseries 111} no.~4, (2025) 044003},
  \href{https://arxiv.org/abs/2408.10352}{{\ttfamily arXiv:2408.10352
  [gr-qc]}}.

\bibitem{Affleck:1983mk}
I.~Affleck, M.~Dine, and N.~Seiberg, ``{Dynamical Supersymmetry Breaking in
  Supersymmetric QCD},''
  \href{https://dx.doi.org/10.1016/0550-3213(84)90058-0}{{\em Nucl. Phys. B}
  {\bfseries 241} (1984) 493--534}.

\bibitem{Horndeski:1974wa}
G.~W. Horndeski, ``{Second-order scalar-tensor field equations in a
  four-dimensional space},'' \href{https://dx.doi.org/10.1007/BF01807638}{{\em
  Int. J. Theor. Phys.} {\bfseries 10} (1974) 363--384}.

\bibitem{Deffayet:2011gz}
C.~Deffayet, X.~Gao, D.~A. Steer, and G.~Zahariade, ``{From k-essence to
  generalised Galileons},''
  \href{https://dx.doi.org/10.1103/PhysRevD.84.064039}{{\em Phys. Rev. D}
  {\bfseries 84} (2011) 064039},
  \href{https://arxiv.org/abs/1103.3260}{{\ttfamily arXiv:1103.3260 [hep-th]}}.

\bibitem{Hui:2012qt}
L.~Hui and A.~Nicolis, ``{No-Hair Theorem for the Galileon},''
  \href{https://dx.doi.org/10.1103/PhysRevLett.110.241104}{{\em Phys. Rev.
  Lett.} {\bfseries 110} (2013) 241104},
  \href{https://arxiv.org/abs/1202.1296}{{\ttfamily arXiv:1202.1296 [hep-th]}}.

\bibitem{DeLuca:2023mio}
V.~De~Luca, J.~Khoury, and S.~S.~C. Wong, ``{Nonlinearities in the tidal Love
  numbers of black holes},''
  \href{https://dx.doi.org/10.1103/PhysRevD.108.024048}{{\em Phys. Rev. D}
  {\bfseries 108} no.~2, (2023) 024048},
  \href{https://arxiv.org/abs/2305.14444}{{\ttfamily arXiv:2305.14444
  [gr-qc]}}.

\bibitem{Santos}
J.~E. Santos. Private communication.

\bibitem{Bekenstein:1972ny}
J.~D. Bekenstein, ``{Transcendence of the law of baryon-number conservation in
  black hole physics},''
  \href{https://dx.doi.org/10.1103/PhysRevLett.28.452}{{\em Phys. Rev. Lett.}
  {\bfseries 28} (1972) 452--455}.

\bibitem{Sotiriou:2013qea}
T.~P. Sotiriou and S.-Y. Zhou, ``{Black hole hair in generalized scalar-tensor
  gravity},'' \href{https://dx.doi.org/10.1103/PhysRevLett.112.251102}{{\em
  Phys. Rev. Lett.} {\bfseries 112} (2014) 251102},
  \href{https://arxiv.org/abs/1312.3622}{{\ttfamily arXiv:1312.3622 [gr-qc]}}.

\bibitem{Babichev:2017guv}
E.~Babichev, C.~Charmousis, and A.~Leh\'ebel, ``{Asymptotically flat black
  holes in Horndeski theory and beyond},''
  \href{https://dx.doi.org/10.1088/1475-7516/2017/04/027}{{\em JCAP} {\bfseries
  04} (2017) 027}, \href{https://arxiv.org/abs/1702.01938}{{\ttfamily
  arXiv:1702.01938 [gr-qc]}}.

\bibitem{Creminelli:2020lxn}
P.~Creminelli, N.~Loayza, F.~Serra, E.~Trincherini, and L.~G. Trombetta,
  ``{Hairy Black-holes in Shift-symmetric Theories},''
  \href{https://dx.doi.org/10.1007/JHEP08(2020)045}{{\em JHEP} {\bfseries 08}
  (2020) 045}, \href{https://arxiv.org/abs/2004.02893}{{\ttfamily
  arXiv:2004.02893 [hep-th]}}.

\bibitem{Hui:2021cpm}
L.~Hui, A.~Podo, L.~Santoni, and E.~Trincherini, ``{Effective Field Theory for
  the perturbations of a slowly rotating black hole},''
  \href{https://dx.doi.org/10.1007/JHEP12(2021)183}{{\em JHEP} {\bfseries 12}
  (2021) 183}, \href{https://arxiv.org/abs/2111.02072}{{\ttfamily
  arXiv:2111.02072 [hep-th]}}.

\bibitem{Chakraborty:2025zyb}
S.~Chakraborty, P.~Heidmann, and P.~Pani, ``{Fermionic Love of Black Holes in
  General Relativity},'' \href{https://arxiv.org/abs/2508.20155}{{\ttfamily
  arXiv:2508.20155 [gr-qc]}}.

\bibitem{Pereniguez:2025jxq}
D.~Pere{\~n}iguez and E.~Karnickis, ``{On the non-zero Love numbers of magnetic
  black holes},'' \href{https://arxiv.org/abs/2509.12418}{{\ttfamily
  arXiv:2509.12418 [gr-qc]}}.

\bibitem{Costa:2015gol}
M.~S. Costa, L.~Greenspan, M.~Oliveira, J.~Penedones, and J.~E. Santos,
  ``{Polarised Black Holes in AdS},''
  \href{https://dx.doi.org/10.1088/0264-9381/33/11/115011}{{\em Class. Quant.
  Grav.} {\bfseries 33} no.~11, (2016) 115011},
  \href{https://arxiv.org/abs/1511.08505}{{\ttfamily arXiv:1511.08505
  [hep-th]}}.

\bibitem{Nair:2024mya}
S.~Nair, S.~Chakraborty, and S.~Sarkar, ``{Asymptotically de Sitter black holes
  have nonzero tidal Love numbers},''
  \href{https://dx.doi.org/10.1103/PhysRevD.109.064025}{{\em Phys. Rev. D}
  {\bfseries 109} no.~6, (2024) 064025},
  \href{https://arxiv.org/abs/2401.06467}{{\ttfamily arXiv:2401.06467
  [gr-qc]}}.

\bibitem{Franzin:2024cah}
E.~Franzin, A.~M. Frassino, and J.~V. Rocha, ``{Tidal Love numbers of static
  black holes in anti-de Sitter},''
  \href{https://dx.doi.org/10.1007/JHEP12(2024)224}{{\em JHEP} {\bfseries 12}
  (2025) 224}, \href{https://arxiv.org/abs/2410.23545}{{\ttfamily
  arXiv:2410.23545 [hep-th]}}.

\bibitem{Cardoso:2018ptl}
V.~Cardoso, M.~Kimura, A.~Maselli, and L.~Senatore, ``{Black Holes in an
  Effective Field Theory Extension of General Relativity},''
  \href{https://dx.doi.org/10.1103/PhysRevLett.121.251105}{{\em Phys. Rev.
  Lett.} {\bfseries 121} no.~25, (2018) 251105},
  \href{https://arxiv.org/abs/1808.08962}{{\ttfamily arXiv:1808.08962
  [gr-qc]}}. [Erratum: Phys.Rev.Lett. 131, 109903 (2023)].

\bibitem{DeLuca:2024nih}
V.~De~Luca, B.~Khek, J.~Khoury, and M.~Trodden, ``{Tidal Love numbers of analog
  black holes},'' \href{https://dx.doi.org/10.1103/PhysRevD.111.044069}{{\em
  Phys. Rev. D} {\bfseries 111} no.~4, (2025) 044069},
  \href{https://arxiv.org/abs/2412.08728}{{\ttfamily arXiv:2412.08728
  [gr-qc]}}.

\bibitem{DeLuca:2022tkm}
V.~De~Luca, J.~Khoury, and S.~S.~C. Wong, ``{Implications of the weak gravity
  conjecture for tidal Love numbers of black holes},''
  \href{https://dx.doi.org/10.1103/PhysRevD.108.044066}{{\em Phys. Rev. D}
  {\bfseries 108} no.~4, (2023) 044066},
  \href{https://arxiv.org/abs/2211.14325}{{\ttfamily arXiv:2211.14325
  [hep-th]}}.

\bibitem{Barbosa:2025uau}
S.~Barbosa, P.~Brax, S.~Fichet, and L.~de~Souza, ``{Running Love numbers and
  the Effective Field Theory of gravity},''
  \href{https://dx.doi.org/10.1088/1475-7516/2025/07/071}{{\em JCAP} {\bfseries
  07} (2025) 071}, \href{https://arxiv.org/abs/2501.18684}{{\ttfamily
  arXiv:2501.18684 [hep-th]}}.

\bibitem{Barura:2024uog}
C.~G.~A. Barura, H.~Kobayashi, S.~Mukohyama, N.~Oshita, K.~Takahashi, and
  V.~Yingcharoenrat, ``{Tidal Love numbers from EFT of black hole perturbations
  with timelike scalar profile},''
  \href{https://dx.doi.org/10.1088/1475-7516/2024/09/001}{{\em JCAP} {\bfseries
  09} (2024) 001}, \href{https://arxiv.org/abs/2405.10813}{{\ttfamily
  arXiv:2405.10813 [gr-qc]}}.

\bibitem{Bhattacharyya:2025slf}
A.~Bhattacharyya, S.~Ghosh, N.~Kumar, S.~Kumar, and S.~Pal, ``{Love beyond
  Einstein: Metric reconstruction and Love number in quadratic gravity using
  WEFT},'' \href{https://arxiv.org/abs/2508.02785}{{\ttfamily arXiv:2508.02785
  [hep-th]}}.

\bibitem{Goldberger:2019sya}
W.~D. Goldberger and I.~Z. Rothstein, ``{An Effective Field Theory of Quantum
  Mechanical Black Hole Horizons},''
  \href{https://dx.doi.org/10.1007/JHEP04(2020)056}{{\em JHEP} {\bfseries 04}
  (2020) 056}, \href{https://arxiv.org/abs/1912.13435}{{\ttfamily
  arXiv:1912.13435 [hep-th]}}.

\bibitem{Saketh:2023bul}
M.~V.~S. Saketh, Z.~Zhou, and M.~M. Ivanov, ``{Dynamical tidal response of Kerr
  black holes from scattering amplitudes},''
  \href{https://dx.doi.org/10.1103/PhysRevD.109.064058}{{\em Phys. Rev. D}
  {\bfseries 109} no.~6, (2024) 064058},
  \href{https://arxiv.org/abs/2307.10391}{{\ttfamily arXiv:2307.10391
  [hep-th]}}.

\bibitem{Mandal:2023hqa}
M.~K. Mandal, P.~Mastrolia, H.~O. Silva, R.~Patil, and J.~Steinhoff,
  ``{Renormalizing Love: tidal effects at the third post-Newtonian order},''
  \href{https://dx.doi.org/10.1007/JHEP02(2024)188}{{\em JHEP} {\bfseries 02}
  (2024) 188}, \href{https://arxiv.org/abs/2308.01865}{{\ttfamily
  arXiv:2308.01865 [hep-th]}}.

\bibitem{Jakobsen:2023pvx}
G.~U. Jakobsen, G.~Mogull, J.~Plefka, and B.~Sauer, ``{Tidal effects and
  renormalization at fourth post-Minkowskian order},''
  \href{https://dx.doi.org/10.1103/PhysRevD.109.L041504}{{\em Phys. Rev. D}
  {\bfseries 109} no.~4, (2024) L041504},
  \href{https://arxiv.org/abs/2312.00719}{{\ttfamily arXiv:2312.00719
  [hep-th]}}.

\end{thebibliography}\endgroup

\onecolumngrid

\appendix

\section*{APPENDICES}

\section{From Einstein-Hilbert to Poincar\'e in the static sector}\label{app:dualization}

We start from the action for time-independent perturbations in the field basis defined by the temporal Kaluza-Klein reduction in eq.~\eqref{eq:temporalKK}. In this section all three-dimensional indices $i,j,k, \dots$ are raised and lowered with the three-dimensional metric $\gamma_{ij}$ and its inverse $\gamma^{ij}$. The Einstein-Hilbert action in the static sector takes the form
\begin{equation}
S_{\rm EH} = - \dfrac{1}{16\pi G} \int dt d^{3}{x} \sqrt{\gamma} \; \left(-R^{(3)}[\gamma] + 2 \partial_i\phi \partial^i \phi - \dfrac{1}{4}e^{4\phi} F_{ij} F^{ij}\right).
\end{equation}
We now introduce a vector field $F_i$ by dualizing the static field strength $F_{ij}$:
\begin{equation}\label{eq:Fdual1}
F_i = \dfrac{1}{2}\epsilon_{ i}^{\phantom{a}jk} F_{jk},
\end{equation}
where $\epsilon_{ijk}$ is the three-dimensional covariant Levi-Civita tensor defined by the metric $\gamma_{ij}$. We have ${2 F_i F^i = F_{ij}F^{ij}}$.
Moreover, because of the Bianchi identity on $F_{jk}$ it follows that $\nabla^i F_i=0$. We can then impose this as a constraint in the action:
\begin{equation}
S_{\rm EH} = - \dfrac{1}{16\pi G} \int dt d^{3}{x} \sqrt{\gamma} \; \left(-R^{(3)}[\gamma] + 2 \partial_i\phi \partial^i\phi - \dfrac{1}{2} e^{4\phi} F_{i} F^{i} - a \nabla^i F_i\right),
\end{equation}
and dualize $F_i$ to a pseudoscalar field $a$. 
To do this we first integrate by parts the last term and then integrate out~$F_i$, which appears only quadratically in the action. We find $F_i =e^{-4\phi} \partial_i a$. Plugging back we obtain
\begin{equation}
S_{\rm EH} = - \dfrac{1}{16\pi G} \int dt d^{3}{x} \sqrt{\gamma} \; \left(-R^{(3)}[\gamma] + 2 \partial_i\phi \partial^i\phi + \dfrac{e^{-4\phi}}{2}\partial_i a \partial^i a \  \right) .
\end{equation}
Notice that, since the Kaluza-Klein decomposition was done on an infinite, rather then compact, time direction, $A_i$ is an $\mathbb{R}$ gauge field rather than ${\rm U}(1)$. As a consequence, the pseudoscalar $a$ is defined on $\mathbb{R}$ and is not periodic.

We can put the $\phi$-kinetic term in a form similar to the $a$-kinetic term by rewriting it as ${e^{-4\phi}}(\partial_i e^{2\phi}) (\partial^i e^{2\phi})/2$.
Introducing the complex field
\begin{equation}
z = a + i\, e^{2\phi},
\end{equation}
we can rewrite the action as
\begin{equation}
S_{\rm EH} = - \dfrac{1}{16\pi G} \int dt d^{3}{x} \sqrt{\gamma} \; \left(-R^{(3)}[\gamma] + \dfrac{1}{2} \dfrac{\partial_i z \,\partial^i\bar{z}}{({\rm Im}\,z)^2}  \
\right),
\end{equation}
where $z$ is defined in the upper-half complex plane ${\rm Im}(z)>0$.

\vspace{10pt}
{\bf The Weyl tensor in $z$ variables.}
The invariant definition of Love numbers is given in terms of worldline operators built from the Weyl tensor. To translate the selection rules to statements on their coefficients it is useful to rewrite the Weyl tensor in terms of the $z$ or $w$ variables, to make manifest their symmetry properties.

We can first rewrite $E$ and $B$, as defined in eq.~\eqref{eq:Weyl}, in terms of $e^{2\phi}$ and $a$. The result is 
\begin{align}\label{eq:Eij}
E_{ij} =& \dfrac{1}{2}\nabla_i\nabla_j e^{2\phi} -\dfrac{1}{2}e^{-2\phi} \nabla_i a \nabla_j a +\dfrac{1}{2} e^{2\phi} R^{(3)}_{ij}[\gamma] - {\rm trace}, \\
B_{ij} =& \dfrac{1}{2} \nabla_i\nabla_j a +\dfrac{1}{4}e^{-2\phi} \nabla_{(i} a \nabla_{j)} e^{2\phi} - {\rm trace},
\end{align}
where $\nabla$ is the covariant derivative of $\gamma_{ij}$.

We now express the result in terms of the $z$ variable.
We arrive at
\begin{align}
B_{ij} + i\, E_{ij} =& \dfrac{1}{2} \nabla_i \nabla_j z + \dfrac{1}{4} \dfrac{1}{z-\bar{z}} \left(2 \nabla_i z \nabla_j z +\nabla_i z \nabla_j \bar{z} +\nabla_i \bar{z} \nabla_j z \right) + i \, \dfrac{{\rm Im} z}{2} R_{ij}[\gamma]^{(3)} - {\rm trace},\\
B_{ij} - i\, E_{ij} =& \dfrac{1}{2} \nabla_i \nabla_j \bar{z} - \dfrac{1}{4} \dfrac{1}{z-\bar{z}} \left(2 \nabla_i \bar{z} \nabla_j \bar{z} +\nabla_i z \nabla_j \bar{z} +\nabla_i \bar{z} \nabla_j z \right) - i \, \dfrac{{\rm Im} z}{2} R_{ij}[\gamma]^{(3)}- {\rm trace}.
\end{align}
We can also rewrite $E$ and $B$ in terms of $w$. At linear order in $w$ and $\gamma$ we find:
\begin{equation}
E_{ij} -i B_{ij} = \partial_i \partial_j w +(\partial^2 \gamma)_{ij}+ \dots ,\quad E_{ij}  + i B_{ij} = \partial_i \partial_j \bar w+(\partial^2 \gamma)_{ij} + \dots \; ,
\end{equation}
where
\begin{equation}
(\partial^2 \gamma)_{ij}=\dfrac{1}{4}\left(-\square \gamma_{ij} +\partial_i \partial_k \gamma^k_j +\partial_j \partial_k \gamma^k_i - \partial_i \partial_j \gamma \right)\, .
\end{equation}

\subsection{Symmetry group structure} \label{app:symmetry}

Writing the action in terms of $z$ and $\gamma_{ij}$, it becomes manifest that the static sector of four dimensional gravity is invariant under a ${\rm PSL}(2,\mathbb{R})$ group, acting on $z$ by fractional linear transformations
\begin{equation}
z \to \dfrac{\a z+\b }{\c z+\d }, \quad {\rm with } \quad \a \d-\b \c =1,
\end{equation}
where the coefficients are real numbers. The symmetry is the projective version of ${\rm SL}(2,\mathbb{R})$, since $-\mathbf{1}$ acts trivially. This is a version of electric-magnetic duality but is realized as a genuine symmetry in the static sector of $D=4$ GR, in the absence of matter or sources. It includes the $T$ and $S$ modular-type transformations
\begin{equation}
T: z \to z+ b, \quad {\rm and} \quad S: z \to -\dfrac{1}{z}.
\end{equation}
The first one corresponds to the continuous shift symmetry of $a$, while the second one is an extension of the $\mathbb{Z}_2$ symmetry $\phi\to -\phi$ enjoyed by the even sector, to fully nonlinear static GR. Notice that $a$ is invariant under arbitrary shifts, rather than discrete ones. As a consequence $T$ and $S$ generate the whole of ${\rm PSL}(2,\mathbb{R})$, rather than the modular group ${\rm PSL}(2,\mathbb{Z})$, as would be the case if $T$ was a discrete shift.
In addition, the theory is invariant under 
\begin{equation}
P: z \to -\bar{z},
\end{equation}
which corresponds to $a(x)\to -a(x)$. This can be identified as a parity transformation.

The ${\rm PSL}(2,\mathbb{R})$ symmetry is realized nonlinearly, and in fact the lagrangian for $z$ can be identified as the nonlinear sigma model (NLSM) for the coset space ${\rm PSL}(2,\mathbb{R})/{\rm SO}(2)$, which corresponds to Euclidean $\rm AdS_2$. This is nothing else than two-dimensional hyperbolic space and indeed the metric of this NLSM is the metric of the Poincar\'e half-plane model of hyperbolic geometry
\begin{equation}
ds^2 = \frac12 \dfrac{dzd\bar{z}}{({\rm Im}\, z)^2}, \qquad \qquad {\rm Im}\, z>0.
\end{equation}
The Minkowski background (the ``vacuum'') corresponds to $a=0$ and $\phi=0$, that is $z=i$.
It can be useful to map the upper-half plane to the unit disk by a conformal map
\begin{equation}
w = \dfrac{z-i}{z+i},
\end{equation}
which sends the vacuum to $w=0$. The metric becomes that of the Poincar\'e disk model
\begin{equation}
S_{\rm EH} = - \dfrac{1}{16\pi G} \int dt d^{3}{x} \sqrt{\gamma} \; \left(-R^{(3)}[\gamma] + 2 \dfrac{\partial_i w \,\partial^i\overline{w}}{(1 -\overline{w} w)^2}  \
\right).
\end{equation}
In this parametrization the ${\rm SO}(2)$ symmetry that is unbroken by the Minkowski vacuum becomes manifest as a rotation around the origin $w \to e^{i \theta} w$. In terms of $z$ this corresponds to
\begin{equation}
z\to \dfrac{\cos\left(\frac{\theta}{2} \right)z+\sin\left(\frac{\theta}{2} \right)}{ 
-\sin\left(\frac{\theta}{2} \right)z+ \cos\left(\frac{\theta}{2} \right)}, \qquad 0\leq \theta < 2\pi \, ,
\end{equation}
which acts linearly on fluctuations around $z=i$. It includes the duality transformation $S: z \to -1/z$, which corresponds to $\theta = \pi$, and gets mapped to $S: w \to -w$. The symmetry $T: z\to z+b $ becomes instead $T: w \to (2 i\, w - b w +b)/(2 i - b w +b)$.  These two transformation can be used to generate the whole ${\rm PSL}(2,\mathbb{R})$.
The symmetry $P: z\to - \bar{z}$, on the other hand, corresponds to $P: w \to \overline{w}$. Combining $S$ and $P$ we also have that $SP: w \to -\overline{w}$ is a nonlinear symmetry of the static sector of gravity. 

In the following, it will be useful to keep in mind the linear expansion of $w$ as a function of $a$ and $\phi$. We have
\begin{equation}\label{wlinear}
w = \phi - i \, \dfrac{a}{2} + \dots 
\end{equation}
At linear order, the symmetries we previously discussed act as $S: \phi \to -\phi$, $T: a \to a + {\rm const}$ and $P: a \to -a$, while leaving the other fields unchanged. This simple transformation rules for $\phi$ and $a$ under an $S$-type transformation, however, are an artifact of the linear approximation. To make full use of the symmetry at nonlinear level, and treat operators involving $B_{ij}$ (and thus $a$) it is much more convenient to work in terms of the $w$ variables.

\vspace{10pt}
{\bf Symmetry structure including worldline couplings.}
We summarize in Table~\ref{tab:symmetry_action} the action of the symmetries on the variables $z$ and $w$, together with the spurionic transformation rule for $M$. 
\begin{table}[H]
\begin{tabular}{c|cc||c}
$\phantom{a}$  & $z$     & $w$ & $M$  \\ \hline \Tstrut
$S$ & $-\dfrac{1}{z} $ &  $-w$   &  $-M$      \\ \Tstrut
$P$ &    $-\bar{z}$     &  $\overline{w}$ &  $+M$     \\ \Tstrut
$SP$ &    $\dfrac{1}{\bar{z}} $     &  $-\overline{w}$   &  $-M$  
\end{tabular}
\caption{Symmetry action of $S,P,SP$ on the field variables $z$ and $w$, together with the spurionic transformation laws of the mass $M$ for the linear worldline couplings of a point-particle.}\label{tab:symmetry_action}
\end{table}

\section{Einstein-Hilbert in $D$ dimensions}\label{app:higherD}

We start considering pure gravity in $D$ dimensions in the mostly plus signature. The Einstein-Hilbert action in the bulk is given by 
\begin{equation}
S_{\rm EH} = - \dfrac{1}{16\pi G} \int d^D{x} \sqrt{-g} (-R).
\end{equation}
To study gravitational perturbations we adopt a temporal Kaluza-Klein decomposition of the metric, as done also in Ref.~\cite{Kol:2011vg}:
\begin{equation}
ds^2 = - e^{2\phi} \left(dt - A_i dx^i \right)^2 + e^{-\frac{2\phi}{D-3}} \gamma_{ij} dx^i dx^j.
\end{equation}
Its inverse is given by
\begin{equation}
    g^{00} = 
    -e^{-2\phi} + e^{\frac{2\phi}{D-3}}A_iA^i\,, 
    \quad 
    g^{0i} = 
    e^{\frac{2\phi}{D-3}}A^i\,, 
    \quad
    g^{ij} = e^{\frac{2\phi}{D-3}} \gamma^{ij} \, .
\end{equation}
With respect to the standard Kaluza-Klein decomposition we have performed some extra Weyl rescalings of the lower dimensional metric to simplify the action for $\phi$ and the background field expansion. 

The full action in the static limit (with all fields independent of time) takes the form
\begin{equation}
S_{\rm EH} = - \dfrac{1}{16\pi G} \int dt d^{D-1}{x} \sqrt{\gamma} \; \left(-R^{(D-1)}[\gamma] + \left(\dfrac{D-2}{D-3}\right) \gamma^{ij} \partial_i\phi \partial_j\phi - \dfrac{1}{4}e^{2\frac{D-2}{D-3}\phi} \gamma^{ik}\gamma^{jl} F_{ij} F_{kl}\right),
\end{equation}
where $F_{ij}= \partial_i A_j - \partial_j A_i $.
This truncation of the field variables is consistent since we will be interested in a static background and at classical level only static intermediate states will be relevant.

As argued in the main text, the $F^2$ operator can be set to zero when analyzing the classical static response of electric and tensor type. In this case the action has a $\mathbb{Z}_2$ symmetry which acts as $\phi \to -\phi$ and leaves $\gamma^{ij}$ invariant. This is an exact symmetry of the static sector of the nonlinear gravitational action $S_{\rm EH}$, valid both in the electric and tensor sectors, as first noticed in~\cite{Kol:2011vg}.

We can now use this symmetry to argue that the static background solution obtained by resumming point-particle insertions has the property that $\bar{\phi}$ and $\bar{\gamma}_{ij}$ are respectively an odd and even function of $X$, or equivalently of $M$. Indeed, for a spherically symmetric solution, the only dimensionless quantity one can construct is $X\propto GM/r^{D-3}$. The symmetry of the action and the fact that the solution can be constructed perturbatively by using symmetric interaction vertices---with $\phi$ coupling linearly to $M$ and $\gamma$ coupling only to $\phi^2$---imply directly that $\bar{\phi}$ and $\bar{\gamma}_{ij}$ are an odd and even function of $X$, respectively.

We can check this by writing the spherically symmetric Schwarzschild--Tangherlini solution in isotropic coordinates, which takes the form
\begin{equation}
ds^2 = -\left(\dfrac{1-X}{1+X} \right)^2 dt^2 + \Big(1+X\Big)^{\frac{4}{D-3}} \delta_{ij}dx^i dx^j,
\end{equation}
where 
\begin{equation}\label{eq:Xiso-D}
X= \dfrac{1}{4}\left(\dfrac{R_S}{r}\right)^{D-3},\quad R_S^{D-3} = 2GM \dfrac{4\pi \Gamma\left(\frac{D-1}{2}\right)}{(D-2)\pi^{(D-1)/2} }, \quad r= \sqrt{{\bf x}^2} \; ,
\end{equation}
and the coordinates cover the exterior of the black hole $r>R_S/4$.

Perturbation theory in the worldline EFT in term of Kaluza-Klein variables is known to reproduce the PM expanded Schwarzschild--Tangherlini background in isotropic coordinates~\cite{Kol:2011vg,Ivanov:2022hlo}, therefore we will work in this coordinate system when finding the classical black hole background in GR and in its EFT extensions.

The $D$-dimensional Schwarzschild--Tangherlini  background is matched for 
\begin{equation}
e^{2\bar{\phi}} = \left(\frac{1-X}{1+X}\right)^2, \quad \bar{\gamma}_{ij} = \Big(1-X^2\Big)^{\frac{2}{D-3}} \delta_{ij} ,
\end{equation}
with $\bar{A_i}=0$. As anticipated, the transformation $\phi \to - \phi$, $M\to -M$ is a symmetry of the background solution, as well as of the gravitational action restricted to $\phi$ and $\gamma$.

\end{document}